\documentclass[a4paper,11pt]{article}
\pdfoutput=1 

\usepackage{jcappub} 

\usepackage[T1]{fontenc} 

\usepackage{subcaption}
\usepackage{siunitx}
\usepackage{hyperref}
\usepackage{braket}

\definecolor{red}{rgb}{0.9, 0,0}
\definecolor{cerulean}{rgb}{0., 0.62,0.9}
\definecolor{navy}{rgb}{0.05, 0.05,0.8}
\definecolor{orange}{rgb}{0.8, 0.4, 0.}

\newcommand{\dd}{\mathrm{d}}
\renewcommand{\eqref}[1]{Eq.~(\ref{#1})}

\definecolor{pastelBrown}{RGB}{181, 150, 114}

\title{The fast, the slow and the merging:\\
 probes of evaporating memory burdened PBHs}

\author[c]{Alessandro Dondarini}
\author[a,b]{Giulio Marino}
\author[a,b]{Paolo Panci}
\author[b]{Michael Zantedeschi}

\affiliation[a]{Dipartimento di Fisica E. Fermi, Universit\`a di Pisa, Largo B. Pontecorvo 3, I-56127 Pisa, Italy}
\affiliation[b]{INFN, Sezione di Pisa, Largo Bruno Pontecorvo 3, I-56127 Pisa, Italy}
\affiliation[c]{Galileo Galilei Institute for Theoretical Physics, Largo Enrico Fermi 2, I-50125 Firenze, Italy}

\emailAdd{alessandro.dondarini@phd.unipi.it}
\emailAdd{giulio.marino@phd.unipi.it}
\emailAdd{paolo.panci@unipi.it}
\emailAdd{michael.zantedeschi@pi.infn.it}

\abstract{
The so-called memory-burden effect implies that evaporating Primordial Black Holes (PBHs) inevitably stabilize before complete decay. This stabilization opens a new mass window for PBH Dark Matter below $10^{15}\,\mathrm{g}$. The transition to the memory-burdened phase is not instantaneous but unfolds over cosmological timescales, with some PBHs entering this phase in the present epoch. Additionally, a fraction of PBHs undergo mergers today, forming “young” semiclassical black holes that evaporate at unsuppressed rates. Both processes generate fluxes of stable astrophysical particles, which are constrained by current measurements of high-energy $\gamma$-rays and neutrinos. Moreover, the steep increase in energy injection at higher redshifts perturbs the ionization history of the Universe, leading to complementary bounds from observations of the CMB temperature and polarization anisotropies. We find that the reopened window enabled by the memory-burden effect is largely within reach of detection, both locally and across cosmological distances. We further describe how our findings restrict the values of the critical exponent characterizing the memory burden phenomenon. 
}

\begin{document}
\maketitle
\flushbottom
\section{Introduction}\label{sec:intro}
The origin of Dark Matter (DM) is still unknown. A well-motivated candidates are PBHs formed in the early Universe~\cite{Zeldovich:1967lct, Hawking:1971ei,Carr:1974nx,Chapline:1975ojl,Carr:1975qj} (for reviews, see e.g.,~\cite{Carr:2020xqk,Green:2020jor,Escriva:2022duf}). As demonstrated by Hawking~\cite{Hawking:1975vcx}, these objects gradually emit quanta of energy $r_{g}^{-1}$, where $r_g$ is the Schwarzschild radius. At higher-energies, the
spectrum is thermal-like and therefore Boltzmann suppressed, leading to a decrease in the Black Hole (BH) mass $M$, according to the rate $\left({\rm d} M/{\rm d} t\right)_{\text{sc}}  \simeq - r_g^{-2} $.

It is commonly assumed that the semiclassical picture is valid throughout the full lifetime of the BH. If this were to be the case, a BH would require $\mathcal{O}(G_{\rm N} M^2)$ emissions in order to appreciably decrease its mass (where $G_{\rm N}= M_{\rm Pl}^{-2}$ is the Newton constant, determined by Planck mass $M_{\rm Pl}$). This leads to a lifetime
\begin{equation}\label{eq:tausc}
	\tau_{\rm SC}\simeq S\,r_{g}\,,
\end{equation}
where we introduced the so-called Bekenstein-Hawking entropy~\cite{Bekenstein:1973ur} 
\begin{equation} \label{entropy}
	S = 2\pi M \,r_g = \frac{1}{\alpha_{\rm gr}}\,,
\end{equation} 
and $\alpha_{\rm gr}= q_*^2/M_{\rm Pl}^2$ is the gravitational coupling determined by the typical momentum transfer $q_*=r_g^{-1}$. This seemingly innocuous assumption has a series of phenomenological consequences: first of all, only PBHs with a mass approximately larger than $10^{15}\,\rm g$ would be sufficiently long-lived to be the DM. Second, the huge energy injected in both the early, and present-day Universe leads to stringent constraints on the abundance of these objects between $10^{10}\,\rm g$ and $10^{17}\,\rm g$~\cite{Carr:2009jm,Carr:2016hva,Carr:2020gox,Arbey:2019vqx,Boccia:2024nly}. 
As a consequence, the viable mass window for which PBHs can compose $\mathcal{O}(1)$ fraction of the DM is traditionally assumed to be between $10^{17}\,\rm g$ to $10^{23}\,\rm g$ in the so-called ``asteroid-mass" window. For higher values of the masses, dynamical constraints apply such as lensing ones (see e.g.,~\cite{Carr:2020gox}).

\medskip
However, the previous discussion, which assumed the reliability of the semiclassical picture throughout the evolution of the black hole, overlooked the potential impact of quantum backreaction.  Recently, it has been argued that the so-called ``memory burden'' effect~\cite{Dvali:2018xpy,Dvali:2018ytn,Dvali:2020wft,Alexandre:2024nuo, Dvali:2024hsb} halts the evaporation process, thereby stabilizing the PBHs against their decay. This effect is universal in all systems with a high capacity to store information as indicated by several numerical and analytical studies~\cite{Dvali:2018xpy,Dvali:2018ytn,Dvali:2020wft,Dvali:2021tez,Dvali:2021bsy,Dvali:2023qlk,Alexandre:2024nuo,Dvali:2024hsb}. The phenomenon is further independently motivated by its prominence in a large class of objects that have the peculiarity of having the maximal entropy compatible with unitarity-saturation, so-called ``saturons"~\cite{Dvali:2020wqi}.  BHs, possessing an entropy-area law,  are a prime example of such objects. However, saturons can also be found outside of gravity, in renormalizable field theories~\cite{Dvali:2019jjw, Dvali:2019ulr, Dvali:2020wqi,  Dvali:2021jto,  Dvali:2021ooc, Dvali:2021rlf,  Dvali:2021tez, Dvali:2021ofp, Dvali:2023qlk}. Remarkably, saturons display the key-essential properties of BHs such as a thermal rate~\cite{Dvali:2021rlf,  Dvali:2021tez}, a notion of semiclassical information horizon \cite{Dvali:2021tez}, a timescale of information retrieval - compatible with the semiclassical time \eqref{eq:tausc}~\cite{Dvali:2020wqi, Dvali:2021rlf,  Dvali:2021tez} - a bound on their maximal angular momentum, given by the entropy, analogous to the extremality condition in BHs~\cite{Dvali:2021ofp,Zantedeschi:2022czs, Dvali:2023qlk}. This offers the possibility of microscopically understanding some of the properties of the BHs without the necessity of a quantum gravity calculation. Moreover, given the universality, it allows to predict new features that are not accessible within the standard semiclassical analysis of BHs: memory burden is one such property. 

\begin{figure}[h!]
    \centering
    \includegraphics[width=.92\linewidth]{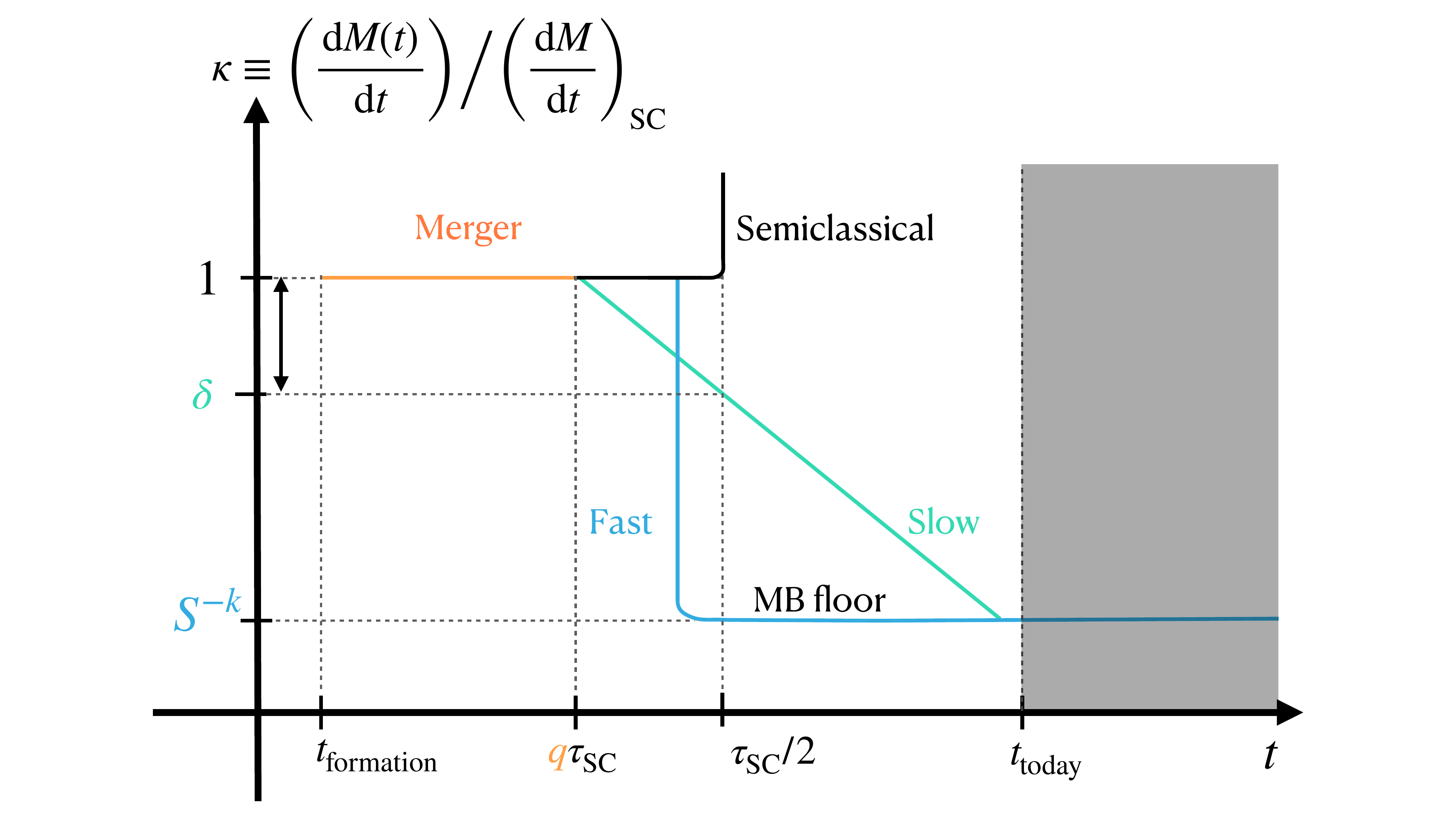}
    \caption{Visual representation of memory burdened BHs evolution. The parameters characterizing the phenomenon are $q$, $\delta$ and $k$ describing the duration of the semiclassical regime, the width of the transition to the memory burdened region, and the emission rate in the memory burdened phase respectively. }
    \label{fig:qualitative}
\end{figure}

When applied to BHs, memory burden has a series of phenomenological implications: in fact, it opens the window for DM for masses below $10^{15}\,\rm g$~\cite{Dvali:2020wft, Alexandre:2024nuo,Thoss:2024hsr,Dvali:2024hsb}. Moreover, it  relaxes the strength of constraints on the lower end of the asteroid-mass window.
The quantity describing the memory burden phenomenon in time, $\kappa(t)$, is the mass-loss rate normalized to the rate of evaporation of the BH at formation time $t_{\rm formation}$ (denoted with the pedix $\rm SC$),
\begin{equation}\label{eq:kappadef}
    \kappa(t) \equiv \left( \frac{{\rm d} M(t)}{{\rm d}t}\right)\Big /  \left( \frac{{\rm d} M}{{\rm d}t}\right)_{\rm SC}\,.
\end{equation}
Incidentally, this quantity also regulates the energy injected over time, making it relevant for phenomenological purposes. The evolution of $\kappa(t)$ is schematically depicted in Fig.~\ref{fig:qualitative} and is described as follows.

The first phase of BH evaporation is purely semiclassical, and consequently, $\kappa\simeq 1$, as indicated by the orange portion of the curve. Integrating the mass loss over this range gives the fraction of energy emitted during this phase, which is captured by the parameter
\begin{equation}\label{qdef}
 q \equiv \Delta M/M\lesssim 1/2\,.
\end{equation} 
If the upper bound in \eqref{qdef} is close to being saturated, the existing semiclassical constraints from Big Bang Nucleosynthesis (BBN), Cosmic Microwave Background (CMB), Galactic and extragalactic $\gamma$-rays~\cite{Carr:2009jm,Carr:2016hva,Carr:2020gox,Arbey:2019vqx} for PBHs of masses between $10^{10}-10^{17}\,$g carry over almost unaltered~\cite{Thoss:2024hsr,Alexandre:2024nuo,Zantedeschi:2024ram}. PBHs of smaller masses are stabilized against their decay before the BBN time thereby avoiding the constraints and, potentially, could then constitute the DM. 
However, it has been noted in~\cite{Zantedeschi:2024ram} that memory burdened PBHs undergo mergers in the present-day Universe, leading to ``young'' BHs. 
Consequently, the resulting PBH emit once more according to the Hawking rate, to then be stabilized,  again, by their memory. Ref.~\cite{Zantedeschi:2024ram} showed that the resulting flux - which is proportional to $q\, \tau_{\rm SC}$ (i.e., the length of the orange line in Fig.~\ref{fig:qualitative}) -  happens to be comparable to present-day astrophysical particle measurements. Such considerations phenomenologically restrict $q\ll 1$. 
Anyway, memory burden can kick in way earlier. In fact, theoretical studies on memory burden suggest $q\gtrsim S^{-1/2}$~\cite{Dvali:2018xpy,Dvali:2018ytn,Dvali:2020wft,Alexandre:2024nuo, Dvali:2024hsb}. Independently motivated studies found the same timescale as the time after which the semiclassical description of BHs is no longer faithful~\cite{Dvali:2011aa,Dvali:2012rt,Dvali:2012en,Dvali:2012wq,Dvali:2013vxa,Dvali:2015wca,Michel:2023ydf}.

As the burden of memory becomes unbearable, $\kappa$ starts decreasing as showed by the cyan curve in  Fig.~\ref{fig:qualitative}. This is quite different as opposed to the semiclassical trajectory showed in black which assumes the BH evolution to be self-similar - meaning that a partially evaporated BH is assumed to be equivalent to a ``young'' BH with the same initial mass (this is also the traditionally assumed trajectory for BH evolution). In such phase, the BH interpolates its rate between the semiclassical one and the memory-burden ``floor'', denoted by the final flat part of the blue curve and given by $S^{-k}$.

Due to the incredibly large change between semiclassical and memory burden rate, it is not surprising that the transition phase, characterized by $\delta$, can take place over cosmological timescales~\cite{Dvali:2025ktz,Montefalcone:2025akm}. Up to logarithmic corrections, $\delta$ describes the mass fraction emitted in the transitioning period. Remarkably, this phase is properly approximated by~\cite{Dvali:2025ktz} \begin{equation}
    \kappa \simeq \frac{\delta\, \tau_{\rm SC}}{2t}, \qquad t\gtrsim \, \tau_{\rm SC}/2\,
\end{equation}
where $t$ is the cosmic time. Due to the steep dependence of the signal as a function of redshift, CMB can significantly constrain the scenario~\cite{Montefalcone:2025akm}. Nevertheless, it is still possible for these objects to source present-day neutrino fluxes~\cite{Dvali:2025ktz} e.g., at \textsc{IceCube}~\cite{IceCube:2018fhm,IceCube:2020wum}. Moreover, the parameter-space window indicated by theory studies~\cite{Dvali:2025ktz} naturally provides a signal of the comparable magnitude with the recent high-energy - larger than $10^2\,$PeV -  neutrino measured by \textsc{KM3NeT} collaboration~\cite{KM3NeT:2025npi}. Several works already tried to link the signal to evaporating PBHs~\cite{Airoldi:2025bgr,Airoldi:2025opo,Anchordoqui:2025xug,Baker:2025cff,Klipfel:2025jql,Jiang:2025blz,Boccia:2025hpm}. In particular, Ref.~\cite{Boccia:2025hpm}, supported by the previous analyses in~\cite{Chianese:2024rsn}, tackled the question of whether the signal could be explained by PBHs in the full memory-burdened phase\footnote{These works may have inadvertently double-counted the neutrino flux by summing both the primary and secondary components extracted from {\tt BlackHawk} v2.3~\cite{Arbey:2019mbc, Arbey:2021mbl}. As clarified in~\cite{Bernal:2022swt}, the secondary flux already includes the primary contribution. Furthermore, as discussed in Appendix~\ref{app:AppBlackHawkclar}, we have identified an issue in the implementation of the HDM option~\cite{Bauer:2020jay} for the hadronization process in {\tt BlackHawk} v2.3. Specifically, the current version seems to account for only half of the fermion flux in the high-energy tail, while including the full contribution in the primary part of the spectrum. Although we have not verified whether this issue is present in earlier versions, it may constitute a potentially relevant source of error in recent studies~\cite{DeRomeri:2024zqs,Zantedeschi:2024ram,Chianese:2024rsn,Boccia:2025hpm,Liu:2025vpz,Jiang:2025blz,Dvali:2025ktz,Chianese:2025wrk,Chaudhuri:2025rcs,Tan:2025vxp,Baker:2025cff,Airoldi:2025bgr,Airoldi:2025opo} that adopt the HDM setting within {\tt BlackHawk}.} which requires a certain degree of fine-tuning in order to achieve the expected flux (notice the extremely sensible suppression appearing in \eqref{tMB} and in \eqref{eq:particleratetime} below).

Obviously, the value of $\delta$ in the transitioning phase is not known and therefore ought to be left arbitrary for the purpose of phenomenological studies. In the limit $\delta\rightarrow 0$, the transitioning phase becomes ``fast'' and is showed by the blue line in Fig.~\ref{fig:qualitative}. So far, most of the phenomenological studies on the memory burden~\cite{Dvali:2021byy,Franciolini:2023osw, Alexandre:2024nuo, Thoss:2024hsr,Dvali:2024hsb,Balaji:2024hpu,Haque:2024eyh,Barman:2024iht,Bhaumik:2024qzd,Barman:2024ufm,Kohri:2024qpd,Borah:2024bcr,Chianese:2024rsn,Zantedeschi:2024ram,Barker:2024mpz,Jiang:2024aju,Loc:2024qbz,Basumatary:2024uwo,Federico:2024fyt,Athron:2024fcj,Barman:2024kfj,Bandyopadhyay:2025ast,Calabrese:2025sfh,Boccia:2025hpm,Liu:2025vpz,Tan:2025vxp}, worked under the simplifying assumption that the transition to the stabilized phase is sharp. We will recap and show why this an assumption justified only in a small corner of parameter space and for specific mass ranges since, in general, PBHs are expected to still be transitioning today.

In this final phase the semiclassical Hawking emission is no longer present. Nevertheless rescattering processes, characterized by integer powers of the gravitational coupling $\alpha_{\rm gr}=S^{-1}$ appearing in~\eqref{entropy}, are still expected. Since $\delta\ll q$, this leads to the approximate lifetime~\cite{Dvali:2020wft,Alexandre:2024nuo,Dvali:2024hsb}, 
\begin{equation} \label{tMB}
	 \tau   \simeq S^{1+k}\, r_g \;,
\end{equation}
   where $k$ is a positive integer parameterizing our ignorance regarding such process (the powers of entropy appearing in \eqref{tMB} can be understood as emerging from the rescattering coupling, determined by $k$, and the $\mathcal{O}(S)$ number of rescattering necessary to change the BH mass by an $\mathcal{O}(1)$ fraction). It follows that $\kappa \simeq S^{-k}$ in this phase. Both numerical and analytic estimates~\cite{Dvali:2020wft,Dvali:2024hsb} suggest $k=2$. For example, $\tau = t_0$ for $k=1,2,3$ leads to $M = 10^7\, \rm g, 3.3 \times 10^3\, \rm g, 3.8 \times 10\, \rm g$ respectively. For the purpose of phenomenological studies, this parameter ought to be left arbitrary.

    In this phase, memory-burdened BHs still emit quanta of energy $\sim 1/r_{g}$, although less often as compared to the semiclassical case, according to the prolonged lifetime \eqref{tMB}. This means that in today's Universe, highly energetic stable astrophysical particles might be sourced by these objects. Refs.~\cite{Thoss:2024hsr,Tan:2025vxp,Chianese:2025wrk} mapped out the constraints due to $\gamma$-rays for different values of $k$. For the $k=2$ case, PBHs lighter than about $10^{5}\,\rm g$ cannot compose the DM due to $\gamma$-rays constraints. Analogous limits emerge due to the emitted neutrino~\cite{Chianese:2024rsn} and electron~\cite{Liu:2025vpz} components. 
    These bounds are derived assuming that through the semiclassical phase the emission is self-similar, namely, that up to the moment of stabilization, the rate of emission is determined by the mass of the PBH in time. However, as pointed in~\cite{Dvali:2025ktz}, this is incorrect. In fact, according to the studies of memory burden, the evaporation cannot track the semiclassical evolution~\cite{Dvali:2018xpy,Dvali:2018ytn,Dvali:2020wft,Dvali:2024hsb}. Therefore, the emission rate is uniquely determined by the initial mass and radius of the BH - both at the level of energetic spectrum and in terms of entropic suppression at the moment of transition to memory burden. Not doing so, introduces a small error which becomes non-negligible if memory burden sets in around $q\lesssim\mathcal{O}(1)$.

    In the present work, we first derive complete constraints from neutrino and $\gamma$-ray measurements, as well as early-Universe physics, in the three relevant phases characterized by: \textit{i)} PBHs evaporating in the full memory burden phase (\textit{fast}); \textit{ii)} PBHs transitioning into the memory burden phase (\textit{slow}); and \textit{iii)} PBHs undergoing mergers that resume Hawking evaporation (\textit{merging}). The first case has already been addressed in the literature~\cite{Thoss:2024hsr,Chianese:2024rsn,Chianese:2025wrk,Liu:2025vpz,Tan:2025vxp}, albeit under the assumption of semiclassical mass tracking, and thus requires reassessment. The latter two phases are novel. For both, only the neutrino component of the signal has been estimated to date~\cite{Zantedeschi:2024ram,Dvali:2025ktz}. To disentangle their distinctive features, we analyze the three scenarios separately, highlighting the regions of parameter space where combined contributions may arise.
    
The rest of this paper is organized as follows. Section~\ref{sec:essence} summarizes the essentials of memory burden and introduces the relevant features for our analysis. Particular focus is invested in clarifying the relation between critical exponents and the parameters that are phenomenologically constrained. In Section~\ref{sec:phenosign}, we compute the neutrino and $\gamma$-ray fluxes for each contribution: the fast, the slow and the merger case and compare them to existing data and sensitivities. Armed with this, in Section~\ref{sec:constr} we derive constraints for the parameters $\delta,q$ and $k$ stemming from indirect probes as well as from the CMB. Afterwards, we summarize our findings as well as the implications regarding the value of the critical exponent. Finally, Section~\ref{sec:conclusion} contains our conclusions.

\section{Memory burden generalities}\label{sec:essence}

\subsection{Prototype Hamiltonian}
As previously stated, memory burden is due to the backreaction of the memory stored in the configuration, which halts its decay. While the phenomenon is universal, we shall 
focus here on a prototype Hamiltonian which captures its essence. For a complete discussion, we refer the interested reader to the original works~\cite{Dvali:2018xpy,Dvali:2018ytn,Dvali:2020wft,Dvali:2024hsb}.

Information is stored in so-called memory modes. We shall label them by an index $j=1,...,K$ denoting their ``flavor''. The total number of ``species'' is denoted by $K$. In the vacuum, their Hamiltonian is given by
\begin{equation}
\label{eq:hfree}
    \hat H_{\rm free} =  \sum_{j=1}^K \epsilon_j\, \hat{n}_{j} \,,
\end{equation}
where $\hat{n}_j \doteq \hat{a}_{j}^\dagger\hat{a}_j$ is the occupation number in terms of creation and annihilation operators obeying canonical commutation relation (CCR). In general, the memory modes can be either fermionic or bosonic. This choice will not alter our conclusions. For definiteness let us proceed with bosonic modes leading to CCR $[\hat{a}_j,\hat{a}_k^{\dagger}] = \delta_{jk}$, $[\hat{a}_j,\hat{a}_k]  =   [\hat{a}_j^{\dagger},\hat{a}_k^{\dagger}] =0$. Without loss of generality, the mass gap of each $j^{th}$ memory mode shall be taken to be the same, $\epsilon_j=\epsilon_{\rm m}$. Similarly, we work, from here onward, with q-bits, $n_j=0,1$.

Let us denote a memory state $|\rm m\rangle$ as 
\begin{equation}\label{eq:memstatesngoldstones}
    |{\rm m}\rangle \doteq |n_1,...,n_K\rangle\,, \quad   N_{\rm G}\doteq\langle {\rm  m}|\sum_j \hat n_j|{\rm m}\rangle\,,
\end{equation}
where the quantity $N_{\rm G}$ counts the total occupation number of q-bits across the flavor space. 

The energy of a given pattern is therefore given by
\begin{equation}\label{eq:memorycost}
    \langle {\rm m}| \hat{H}_{\rm free}|{\rm m}\rangle = \epsilon_{\rm m}\, N_{G}\,.
\end{equation}
A system is very efficient at storing information if a large density of states $n_{\rm st}$ with different memories $|m\rangle, |m'\rangle ,...,$ can fit into a certain small energy gap $\Delta E$. This quantity is set by the physical properties of the system, such as its inverse size. 

If the gap $\Delta E$ is sufficiently small, the different memory patterns provide a large microstate degeneracy which can, in turn, provide a large entropy
    $S= \log n_{\rm st}$ for the configuration.

This is achieved by the second set of modes relevant in our discussion - the master mode $\hat{a}_0,\hat{a}_0^{\dagger}$ which corresponds to bosonic mode obeying CCR. In general, there can be multiple master modes but for our discussion one suffices. In particular, the master mode can assist the memory modes in becoming gapless. In terms of master mode occupation number $\hat n_0\doteq \hat a_0^{\dagger}\hat a_0$, the simplest Hamiltonian - originally proposed in~\cite{Dvali:2017nis, Dvali:2018vvx,Dvali:2018tqi} to describe the mechanism of assisted gaplessness - is 
\begin{equation}\label{eq:htot}
    \hat H = \epsilon_0 \,\hat{n}_0 + \left(1 - \frac{\hat n_0}{N_c} \right)^p \epsilon_{\rm m}\sum_{j=1}^K \hat n_{j}\,,
\end{equation}
where $\epsilon_0$ denotes the mass gap of the master mode. The quantity $N_{c}$ indicates the critical occupation number for which the memory modes become gapless. The Hamiltonian (\ref{eq:htot}) was then adopted in~\cite{Dvali:2018xpy,Dvali:2020wft, Dvali:2024hsb} to discuss memory burden. 

The effective gap of the memory modes is
\begin{equation}\label{eq:effmemgap}
    \epsilon_{\rm m}^{(\rm eff)} = \epsilon_{\rm m}\left(1 - \frac{\hat{n}_0}{N_c} \right)^p\,,
\end{equation}
which is zero for $\langle \hat n_0\rangle = N_c$.
Finally, $p$ is a number $>1$ which plays the role of critical exponent. In fact, it characterizes the shape of the Hamiltonian near the point of memory-modes gaplessness. 

Notice that near the critical region, the microstate degeneracy is given by $n_{\rm st}\simeq 2^{K}$, where for simplicity we consider the case of q-bits, i.e., modes have either occupation number zero or one. This leads to the entropy
    $S \simeq K$ 
meaning that the dimensionality of the flavor space is of order of the BH entropy. 

Lets us denote the full state as $| n_0;{\rm m}\rangle$, where the first entry denotes the occupation number of the master mode, while $\ket{\rm m}$ is defined in \eqref{eq:memstatesngoldstones}. For $ n_0\simeq 0$, the cost of memory is given by \eqref{eq:memorycost}. Near the critical point it is given by
\begin{equation}\label{eq:engapless}
    \langle N_c;{\rm m}|\hat H|{\rm m};N_c\rangle = \epsilon_0 N_c\,.
\end{equation}
This energy is significantly smaller than the one of a memory mode pattern in the vacuum c.f., \eqref{eq:memorycost}. What it implies, is that the region near the critical point acts as a minimum of the Hamiltonian. 

Equivalently, there is an energy barrier due to the impossibility of the system to release memory that resists against any process leading to the decay of the configuration through the emission of $n_0$ quanta. Therefore, Hawking radiation, which tries to interpolate the system between $n_0\simeq S$ to $n_0\sim 0$, necessarily encounters this energy barrier at some point and is stopped. At the level of fundamental Hamiltonian \eqref{eq:htot}, Hawking radiation is achieved through the inclusion of extra interactions terms which we dropped here for simplicity. We shall not deal with those here and refer the interested reader to the relevant literature~\cite{Dvali:2018xpy,Dvali:2020wft, Dvali:2024hsb}.

\subsection{Mapping to black holes}

We now turn to the question as to how to map the above Hamiltonian to BHs. 
For classical BH, the information has zero energy. This is achieved at the level of prototype Hamiltonian \eqref{eq:htot} by having an occupation number $\langle\hat n_{0}\rangle \simeq N_{c}$. For this choice, the energy of the system, c.f., \eqref{eq:engapless}, is
\begin{equation}
\label{eq:classicalen}
   \epsilon_0 N_c = M\,. 
\end{equation}
Hawking radiation corresponds to depletion of $\hat n_{0}$ modes into quanta of energy $1/r_g$; the memory modes are not emitted by Hawking processes which are insensitive to them. Therefore, we can identify $\epsilon_{0}= r_g^{-1}$ and, from \eqref{eq:classicalen}, we obtain $N_{c}= S = r_g^2/G_{\rm N}$. Notice that for $N_{c}\simeq S$, the coupling between memory and master modes in \eqref{eq:htot} is given by $1/S \simeq \alpha_{\rm gr}$. 

\smallskip
A natural question is the origin of the memory modes in BHs. 
As discussed in~\cite{Dvali:2017nis, Dvali:2018xpy,Dvali:2024hsb}, there are  natural - and perhaps unique - candidates. These correspond to the spherical harmonics of the graviton, $Y_{\ell m}$. Validity of the field theory description implies that only harmonics up to the energy cutoff of the theory, $M_{\rm Pl}$, ought to be included, leading to the multiplicity $l^2\simeq   (r_g \, M_{\rm Pl})^2 = S$~\cite{Dvali:2024hsb},
which precisely matches the expected area-law entropy.
It is worth noting that this outcome aligns well with the independent description of gapless modes tied to BHs symmetries presented in~\cite{Averin:2016ybl, Averin:2016hhm}. 

Naturally, most of the modes originate from the highest spherical harmonics. Hence, we identify the BHs’s gapless memory modes with those whose angular momenta scale as $M_{\rm Pl}\,r_g$. Their counterparts in the asymptotic vacuum correspond to the same angular harmonics $Y_{\ell m}$ of a free graviton, but in that region, they possess energy gaps $\epsilon_{\rm m} \simeq M_{\rm Pl}$. This justifies our choice of approximating the memory sector with one and the same energy gap in \eqref{eq:hfree} and clarifies why BHs cannot readily emit information~\cite{Dvali:2018xpy}: for a memory mode to leave a BH through a rescattering process, it must surmount a very large energy barrier, of order $M_{\rm Pl}$, hardly achievable through its soft semiclassical Hawking emission.   

It is not surprising that modes described by the same $Y_{\ell m}$ have significantly different energy gaps inside and outside the BHs. Because the BHs breaks Poincar\'e symmetry at the scale $M_{\rm Pl}$~\cite{Dvali:2020wqi,Dvali:2024hsb}, the BHs’s memory modes remain gapless despite having high orbital angular momentum, whereas the corresponding asymptotic modes acquire gaps on the order of $M_{\rm Pl}$.
These points provide sufficient groundwork for adapting the general Hamiltonian~\eqref{eq:htot} to the BHs setting. Concretely, the effective Hamiltonian governing BHs memory and master modes takes the form of~\eqref{eq:htot}, with the understanding that the index $j$ labels the spherical harmonics numbers~\cite{Dvali:2024hsb}. 

The last ingredient we need is the typical number of occupied q-bits, $N_{G}$, defined in \eqref{eq:memstatesngoldstones}. As discussed in~\cite{Dvali:2024hsb}, $N_{\rm G}$ is determined statistically at the time of formation of the BH. Since different memory patterns lead to the same classical BH it is expected that the probability distribution $\mathcal{P}(N_{G})$ is binomial,
\begin{equation}\label{eq:probbinomial}
    \mathcal{P}(N_G) = \, 2^{-K} \frac{K!}{(K-N_{\rm G})! N_{\rm G}!} \,, 
\end{equation}
implying that the average $\overline{N}_{\rm G}=K/2= S/2$ with a width given by $\sim\sqrt{S}$~\cite{Dvali:2024hsb}.

The mapping of the parameters of the fundamental Hamiltonian to a BH can be summarized as follows
\begin{equation}
    \epsilon_0=r_{g}^{-1}\,,\quad N_{c} = K =S\,,\quad \epsilon_{\rm m}= M_{\rm Pl}\,,\,\quad N_{\rm G}=S/2\,.
\end{equation}
Armed with this, it is straightforward to  
find the point at which $n_0^{(\rm stab)}$, decreasing due to Hawking emissions, gets stabilized by the memory
\begin{equation}
\label{eq:deltanphi}
    \Delta n_{0} = S - n_{0}^{(\rm stab)} = \left(\frac{ 2 }{p\,\sqrt{S}} \right)^{\frac{1}{-1+p} }S\,,
\end{equation}
where we remind that $n_0=S$ corresponds to the initial occupation number of the master modes. 

Since every emission requires a time $r_g$, we can estimate the time-scale of onset of memory burden. For example, for $p=2$ and $p\gg 1$ we have, respectively
\begin{equation}\label{eq:relationpmemory}
    \begin{split}
        &p = 2 \ \qquad \rightarrow \qquad t_{\rm memory} \simeq \sqrt{S}\,r_{g} \,,\\
        &p\gg1 \qquad \rightarrow\qquad t_{\rm memory} \simeq S\, r_g\simeq \tau_{\rm SC}\,.
    \end{split}
\end{equation}
Moreover, the fractional mass emitted during this period is
\begin{equation}
\label{eq:deltamovm}
   q= \frac{\Delta M}{M} \simeq\frac{\Delta n_{0}}{S} \simeq  \left(\frac{2}{p \,\sqrt{S }} \right)^{\frac{1}{-1+p} }\,,
\end{equation}
where $\Delta M =M(t)-M_0$ is the difference between the mass of the BHs and its initial value $M_0$.
For $p=2$, $q\simeq 1/\sqrt{S}$, while for $p\gg 1$, $q\sim \mathcal{O}(1)$. 

\subsection{Memory burden universality}\label{subsec:universality}

The memory burden phenomenon is not unique to BHs but is a generic feature of systems with large information storage capacity. It also manifests within renormalizable field theories without gravity, as explicitly shown by non-topological solitons~\cite{Dvali:2024hsb}. These solitonic bubbles---reminiscent of Q-balls with a global $U(1)$ symmetry~\cite{Friedberg:1976me,Coleman:1985ki,Kusenko:1997vi}---arise in an $SU(N)$ symmetric theory in which the order parameter, a scalar field in the $(N^2 - 1)$ representation, spontaneously breaks the symmetry to $SU(N-1) \times U(1)$ inside the bubble. As a result, a $2(N - 1)$-dimensional flavor space of quasi-Goldstone bosons becomes localized within the bubble. These modes are energetically expensive outside the soliton, due to symmetry restoration, and thus cannot be easily emitted. Consequently, the bubble effectively possesses an ``information horizon'' in the Goldstone flavor space.

At large $N$, one can straightforwardly derive the effective action for the Goldstone modes in the bubble background~\cite{Dvali:2021tez}. In this semiclassical limit, the radial mode couples to the total flavor sum of the Goldstone kinetic terms. Crucially, the macroscopic properties of the bubble are insensitive to the specific occupation pattern of the Goldstone flavor modes. This degeneracy in microstate configurations leads to a large entropy. Unitarity bounds this entropy from above, with the maximal value scaling as the area of the bubble in units of the Goldstone decay constant~\cite{Dvali:2021tez,Dvali:2024hsb}. For BHs, this constant is $M_\mathrm{Pl}$, thereby recovering the Bekenstein area law~\cite{Bekenstein:1973ur}.

This seemingly different system maps naturally to the prototype Hamiltonian~\eqref{eq:htot}. The radial mode corresponds to the master mode $n_0$, spontaneously breaking the symmetry in the bubble interior. The localized Goldstone modes, responsible for the large microstate entropy, play the role of memory modes. Near the critical point $n_0 \lesssim N_c$ (corresponding to the bubble interior), the system is characterized by a critical exponent $p = 3/2$. A key difference from the BHs case is that the memory mode mass is set by the inverse bubble radius, $\epsilon_\mathrm{m} \sim 1/R_\mathrm{bubble}$, realizing a type I memory burden, as opposed to the type II scenario typical of BHs~\cite{Dvali:2024hsb}.
Initially, the soliton exists in a configuration where the master mode dominates the energy budget. As it attempts to collapse, this process is halted by the backreaction from the trapped Goldstone modes. This has been analytically and numerically demonstrated in~\cite{Dvali:2024hsb} (see also the visual summary at the following \href{https://www.youtube.com/watch?v=boDpRXJnT5E&t=75s&ab_channel=MichaelZantedeschi}{URL}). The physical reason is simple: the Goldstone modes are energetically trapped inside the bubble and cannot be efficiently released, forcing a dynamical backreaction on the radial mode that halts further collapse. This stabilization occurs regardless of additional structural features such as bubble wall width or vorticity.

The mechanism just described requires only a few basic ingredients: spontaneous symmetry breaking and the localization of a large flavor space of Goldstone modes, leading to microstate degeneracy and large entropy. These are precisely the essential ingredients described by the prototype Hamiltonian~\eqref{eq:htot}, which explains why the two systems can be mapped onto one another. These features are expected to appear in a wide class of systems capable of storing large amounts of information, in particular those obeying an area law for entropy---so-called \emph{saturons}~\cite{Dvali:2019jjw,Dvali:2019ulr,Dvali:2020wqi,Dvali:2021jto,Dvali:2021ooc,Dvali:2021rlf,Dvali:2021tez,Dvali:2021ofp,Dvali:2023qlk}---of which BHs are a prime example.

In the case of solitonic bubbles, the critical exponent $p$ can be derived from first principles, starting from the microscopic Lagrangian. This exponent characterizes the system's behavior near the gapless point $n_0 \simeq N_c$ and governs the onset of memory burden. This is entirely analogous to critical phenomena near phase transitions, where the dynamics are dominated by universal behavior. BHs, too, are believed to lie near criticality~\cite{Dvali:2011aa} - on the verge of a quantum phase transition (with $\alpha_\mathrm{gr}\, S \simeq \mathcal{O}(1)$) - and thus should likewise be governed by critical exponents. However, due to the absence of a known microscopic theory, the value of $p$ for BHs remains undetermined. In this work, we remain agnostic about its precise value and instead focus on phenomenological constraints on the available memory burden parameter space.

\subsection{Decay in the burdened phase - the fast}
After a BHs enters the memory burden phase, a more detailed analysis is required to understand what happens next. As discussed in~\cite{Dvali:2020wft,Dvali:2024hsb}, two main scenarios arise. The first posits that a new classical (collective) instability develops, causing the (former) BHs to evolve under this instability. Current knowledge does not rule out the possibility that, due to this mechanism, the BHs remnant might disintegrate via a non-linear process. The second scenario assumes no immediate classical instability. This outcome is more conservative since it is suggested by both analytic and numerical studies of the prototype Hamiltonian~\cite{Dvali:2020wft}. It is further observed in the case of memory burdened bubbles~\cite{Dvali:2024hsb}. In this case, the BHs continues to decay quantum-mechanically, but the memory burden makes the process exceedingly slow. As shown in~\cite{Dvali:2020wft,Dvali:2024hsb}, its remaining lifetime is given by~\eqref{tMB}.

 This form reflects the fact that the extended lifetime must be analytic in $S$, because the decay rate itself is analytic in both occupation numbers and gravitational couplings, all of which are determined by $S$. When $k=0$ (i.e., zero memory burden), one recovers the usual Hawking decay rate. The reason for $k>0$ is that the BHs must rid itself of its memory burden in order to keep decaying. In other words, the excited memory modes must be de-excited through scattering processes that involve at least pairs of such modes. Each mode $Y_{\ell m}$ needs a partner $Y_{\ell 'm'}$ with very close values of $\ell,m$ so they can annihilate into modes of lower angular momenta to match their energies. These pairings are extremely rare, with an annihilation rate
$\Gamma \sim \left(\epsilon_{\rm m}^{(\rm eff)}\right)^5/M_{\rm Pl}^4 \sim 1/(r_g^5 M_{\rm Pl}^4).
$~\cite{Dvali:2024hsb}.
Expressed in terms of the original BHs entropy, this leads to a lifetime
$
\tau \simeq r_g\, S^{3},
$
which is consistent with \eqref{tMB} for $k=2$. Notice that the timescale in~\eqref{tMB} might not correspond to the full evaporation of the BH. In fact, $\tau$ is determined by the typical rescattering time of the constituents within the memory burden ``floor.'' As the BH sheds an  $\mathcal{O}(1)$ fraction of its mass, a different process could become dominant, potentially slowing down the evaporation even further. Alternatively, some instability could develop, effectively leading to the complete evaporation of the object.

Of course, from phenomenological point of view, $k$ has to be scanned as a free parameter as done already in several works~\cite{Thoss:2024hsr,Chianese:2024rsn,Liu:2025vpz,Chianese:2025wrk,Tan:2025vxp}. However, these works also assume that in the semiclassical phase the evaporation rate tracks the instantaneous radius of the BH. Due to the smallness of the gravitational coupling, given by the inverse entropy, it is not possible for the system to do so. In fact, the gravitational coupling itself is not expected to evolve in time. Notice that this is precisely tied to the essence of the memory burden effect: the inability of the system to keep up with the changing background. Assuming tracking throughout the semiclassical phase introduces an error when mapping the resulting constraint to the mass of the DM. Such error is clearly not negligible unless $q\ll1$.

As the memory modes are emitted, the burden of the configuration slowly decreases, therefore allowing for the further release of the master mode as well. It is natural to ask how to characterize the emission throughout this quantum phase. We expect the BH to still emit at energy given by its initial radius, although with a rate suppressed by powers of the gravitational coupling. Therefore,
in the full burden phase we have 
    $\kappa = S^{-k}$.
A natural question is how fast the memory burden phase is realized: notice that $\kappa$ becomes extremely small in the full memory burden phase. However, the rate of emission is determined by the same quantity we are trying to suppressed, $\kappa$ itself. Since the energy per emission is constant the transition cannot be instantaneous. This issue has been tackled in~\cite{Dvali:2025ktz} and shall be summed up in the next subsection. 

\subsection{Slow onset of memory burden - the slow}
\label{sec:theslow}

In~\cite{Dvali:2025ktz}, it was pointed out that the transition to the memory burden phase can be slow, potentially resulting in PBHs that are still transitioning today. We briefly recall the salient features of this transition.
The relevant quantity governing the slowdown is the effective gap of the master mode. For a ``young'' BHs, this gap is simply $\epsilon_0 = r_g^{-1}$. However, as the BHs approaches the memory burden phase, the gap is modified by the increasing energy stored in the memory modes. The growing parameter characterizing the hardening of the master mode during this onset is
$\Delta \epsilon_0=p \,M_{\rm Pl} \left(1 - n_0/S \right)^{p-1}/2$. 
This leads to an exponential suppression of emission due to a mismatch between the asymptotic emission energy $r_g^{-1}$ and the master mode gap. The ratio between these two scales is given by~\cite{Dvali:2025ktz}
\begin{equation}
\Delta N =\Delta \epsilon_0 r_g= \frac{p \sqrt{S}}{2} \left( \frac{M_{0} - M(t)}{M_0} \right)^{p-1} \ , 
\end{equation}
where $M_0$ is the initial mass of the BHs.

As a result, the mass-loss rate is exponentially suppressed:
\begin{equation}\label{eq:suppresion_eq}
\frac{{\rm d}M}{{\rm d}t} = \left(\frac{{\rm d}M}{{\rm d}t}\right)_{\rm SC} \left(\frac{1}{S}\right)^{\Delta N} \ .
\end{equation}
A small integer value of $\Delta N$ suffices to suppress the semiclassical rate significantly. For instance, $\Delta N \simeq 2$ already yields a suppression factor of $S^{-2}$, corresponding to a memory burden phase with $k = 2$.

Ref.~\cite{Dvali:2025ktz} addressed the dynamical evolution of PBHs transitioning into memory burden. It was shown that for times $t \gtrsim \tau_{\rm SC}/2$, corresponding to the half-lifetime of the would-be semiclassical evaporation, the transition is well-approximated by
\begin{equation}\label{kappafullmem}
\kappa \simeq \delta \frac{\tau_{\rm SC}}{2t} \ ,
\end{equation}
where $\delta$ roughly characterizes the width of the transition region, i.e., the fraction of mass emitted during this phase. Qualitatively, approximation \eqref{kappafullmem} follows from the observation that increasing exponential suppression of the rate prolongs the time between emissions, thereby stretching the transition. This leads to a logarithmic evolution of the mass. Comparison between the numerical solution to \eqref{eq:suppresion_eq} and the analytic approximation \eqref{kappafullmem} shows perfect agreement for PBHs in the mass-window of interest. We shall therefore adopt \eqref{kappafullmem} in the phenomenological analysis to derive constraints.  

Consistency of the picture requires $\delta \ll q$, since the energy emitted during the transition cannot exceed that emitted in the semiclassical phase - notice that, in the mass window between $10\,\rm g$ and $10^{15}\,\rm g$ 
relevant for this work, $\delta\lesssim 10^{-2}$ ensures that the mass of the PBHs is unaltered over cosmological timescales. Analysis of the prototype Hamiltonian~\eqref{eq:htot} also suggests that $\delta$ is not expected to be much smaller than $q$~\cite{Dvali:2025ktz}. It is, in fact, related to the critical exponent $p$ via
\begin{equation}\label{eq:deltapmapping}
\delta \simeq \frac{2}{(p-1)\ln(S)} S^{\frac{1}{2 - 2p}} \ ,
\end{equation}
which, up to logarithmic and numerical factors depending on $p$, is of order $q$.
As expected, the onset and characteristics of the transition are fully determined by the critical exponent $p$.
However, this result is based on a toy model analysis. Additional terms in~\eqref{eq:htot} could modify the conclusions. Moreover, we worked under the assumption of a single master mode. More in general, there could be multiple ones~\cite{Dvali:2020wft,Dvali:2024hsb,Dvali:2025ktz}. For these reasons, we treat $\delta$ as a phenomenological parameter in the analysis and will comment on the consequences for $p$ afterwards.

If today’s DM consists of PBHs already in the memory burden phase, then $\kappa = S^{-k}$ as in~\eqref{tMB}, and the associated flux can rival current astrophysical backgrounds in certain mass ranges~\cite{Thoss:2024hsr,Chianese:2024rsn}. For instance, with $k = 2$, this holds for PBHs with masses $\lesssim 10^5\,\rm g$. The sensitivity to the value of $k$ is exponential making it less appealing than the case of a slow transition, polynomially determined by $\delta$.

Whether a PBH is in the full memory burden phase at a given time depends on both $\delta$ and $k$. Specifically, $\kappa$ is given by
\begin{equation}
\kappa = \max\left(\frac{\delta \tau_{\rm SC}}{2t},S^{-k}\right).
\end{equation}
Since no \emph{a priori} relation exists between $k$ ($k$ is anyway not expected to be $\gg 1$) and $\delta$, we analyze two limiting cases separately: \textit{i)} PBHs still undergoing transition over cosmological timescales, and \textit{ii)} PBHs that transitioned essentially instantaneously. For certain parameter ranges, both scenarios may be realized within the lifetime of the Universe. We will comment further on this in the analysis below.

\subsection{Merger of memory burdened PBHs}\label{subsec:mergers}

So far we have discussed potential signatures of memory burden BHs stemming from their quantum phase. As previously mentioned, the final constraints rely on the assumption of democracy of the gravitational emission in the Standard Model (SM) species even though the BHs posses macroscopic quantum hair. 

In this regard~\cite{Zantedeschi:2024ram} suggested a way of constraining the scenario that is independent of the width of the transition to the memory burden $\delta$ as well as on the rate of emission in the full memory burdened phase, characterized by $k$. In particular, it was realized that if these objects constitute the dark  matter they undergo mergers in today's Universe, leading to ``young'' BHs thereby emitting with unsuppressed Hawking rate before being stabilized by the memory burden once again. 

To see that this is the case, let us consider, for the sake of argument, two merging BHs of equal mass $M_{\rm PBH}$ (of entropy $S$) stabilized by their burden (or in the process of transitioning to it). The gravitational field away from memory burdened BH is unchanged from the semiclassical case as most of the energy of the object is characterized by the master modes, of energy $r_g^{-1}$. Therefore, the inspiraling phase proceeds analogously to the case of classical merging BHs. This makes the merger inevitable.

A natural question is what happens to the memories of the two progenitors. Notice that the final BH, of roughly twice the original mass, has an area four times larger ($4S$) than the one of progenitor. Therefore, it has more than enough memory to store the full initial information of the progenitors, regardless of their amount of memory, of order $2S$. Furthermore, we cannot exclude that a fraction of the memory is emitted at the merger time, when the BHs quantum hair become relevant, potentially leading to some backreaction on the dynamics. Notice that anyway, the amount of energy available in the memory sector to backreact on the merger dynamics is not large enough to prevent the merger from happening. All in all, it is therefore more than justified to assume that the memory of the newly resulting BH is still determined statistically according to the binomial probability distribution \eqref{eq:probbinomial}. In particular, this ensures that the BH decays semiclassical up to its stabilization.

This point of view is also justified by~\cite{Dvali:2023qlk}. Therein, the merger of BH prototypes in the form of solitonic bubbles stabilized by their memory was addressed. Incidentally, the same system discussed in \cite{Dvali:2024hsb} and recapped Subsec.~\ref{subsec:universality} was adopted. 
The merger dynamics of two such bubble proceeds unaltered until the cores overlap. At that moment, the pre-existing memory of the two progenitor bubbles start interacting, backreacting on the emitted signal at the merger time. In some region of parameter space, this can lead to macroscopic deviations in the ringdown signal - potentially accompanied by the emission of some charge (memory). The resulting bubble is composed of an excited radial mode whose subsequent collapse - akin to Hawking evaporation in the BH counterpart - is stabilized by the residual global charge (memory) inside the bubble.  

To estimate the flux stemming from PBHs merging and resuming their Hawking evaporation, we shall follow~\cite{Zantedeschi:2024ram} and consider solely the contribution due to binary formed in the early Universe. In fact, the merger rate for such light PBHs is primarily sourced by the distribution of PBHs that decouples from the Hubble flow before matter-radiation equality~\cite{Ali-Haimoud:2017rtz, Raidal:2018bbj,Liu:2018ess}, and, for a monochromatic PBH mass distribution, reduces to \cite{Vaskonen:2019jpv, Hutsi:2020sol, Jedamzik:2020ypm, Young:2020scc, Jedamzik:2020omx,Raidal:2024bmm} 
\begin{equation}
    \label{eq:mergerrate}
    R_{\rm PBH}(t) =\frac{5.69\times 10^{-66}}{\rm cm^3\,s}\,f_{\rm PBH}^{\frac{53}{37}}\,\left(\frac{t_0}{t}\right)^{\frac{34}{37}}\,\left(\frac{2 M_{\rm PBH}}{10^{10}\rm{g}} \right)^{-\frac{32}{37}}S_1 \times S_2\,.
\end{equation}
Here $M_{\rm PBH}$ denotes the mass of the PBH composing the DM and the factor $2$ accounts for the final mass post merger. The suppression factors $S_{1}\times S_{2}$ parametrize  two components: for $f_{\rm PBH}\gtrsim 10^{-3}$ (corresponding roughly to the strongest constraints we will obtain below) the term $S_1 \approx 0.24$ is redshift independent and accounts for the interactions between the binary system and the surrounding DM inhomogeneities, as well as neighboring BHs~\cite{Hutsi:2020sol} (we refer the interested reader to the above literature for the precise definition). The second suppression factor, $S_2(y) \approx \min \left[ 1, 9.6 \times 10^{-3} y^{-0.65} \exp \left( 0.03 \ln^2 y \right) \right]$, parametrizes the suppression caused by BHs absorbed by collapsed PBH clusters. Here, $y\equiv \left( t / t_0 \right)^{0.44} f_{\mathrm{PBH}}$. The function $S_2(y)$ attains a minimum at $10^{-2}$, consistently with numerical simulations~\cite{Inman:2019wvr, Franciolini:2022htd}. For light PBHs, dynamical captures induced by gravitational waves as well as late-time dynamical capture are negligible \cite{Franciolini:2022htd}. In~\cite{Franciolini:2022htd} it is further argued that \eqref{eq:mergerrate} can be applied to PBHs in the asteroid-mass window. Although the PBHs considered here are even lighter, we notice that their arguments, showing that potential corrections have a very mild scaling as a function of mass, proceeds unaltered. Notice that Ref.~\cite{Kohri:2024qpd}, which studied the gravitational wave signal due to the inspiraling of memory burdened BHs, adopts also a similar rate following~\cite{Sasaki:2018dmp}. However, the suppression factor is not taken into account to compute the corresponding $\Omega_{\rm GW}$. Furthermore, local non-Gaussianity in primordial curvature perturbations can cluster PBHs at formation, potentially enhancing the merger rate by up to $\mathcal{O}(10^7)$~\cite{Raidal:2017mfl, DeLuca:2021hde, Franciolini:2022htd}. In order to be as conservative as possible, we stick with \eqref{eq:mergerrate} in our analysis. 

Before moving to the next subsection, a comment regarding the spin is in order. After the merger, the resulting BH is expected to have an initial spin of order $ a \sim 0.7 $ (see, e.g.,~\cite{DeLuca:2020bjf}), where $ a $ is the dimensionless spin parameter. Rotation significantly affects the Hawking spectrum on timescales of order $\lesssim 10^{-1}\, \tau_{\rm SC} $~\cite{Page:1976ki,Page:1976df}, which is roughly the time required for the PBH to relax toward a non-rotating state. Therefore, as long as $ q \lesssim \mathcal{O}(1)$, the impact of spin on the total spectrum is expected to be sufficiently diluted, justifying the use of constraints based on a non-spinning BH.
However, we will also derive constraints on the parameter $q $, which, in the mass window of interest, turns out to be $ \lesssim 10^{-2} $ in order to account for the entirety of the DM. In this region of parameter space, spin effects may no longer be negligible. However, we cannot exclude the possibility that memory burden effects influence the merger dynamics, potentially biasing the resulting spin (note that the initial spins of the merging PBHs also contribute to this quantity).
Additionally, the implementation of {\tt BlackHawk} for highly spinning black holes introduces artificial features~\cite{Arbey:2025dnc}. For these reasons, we neglect spin effects in our analysis, noting that their inclusion would only strengthen the resulting constraints.

\subsection{Summary of constrained parameters}

In Fig.~\ref{fig:qualitative} we show a cartooned representation of the time evolution of the parameter $\kappa$ due to the various facets of memory burden we will constrain in the next sections. We summarize them here in order to ease the reader's pain. 

\paragraph{The merger:} In the first phase, the BH evolves according to semiclassical rate and $\kappa \simeq1$. The duration of this period ends when the memory burden kicks in
\begin{equation}
\label{eq:tauq}
    \tau_{\rm MB} \simeq q \,\tau_{\rm SC}\,,
\end{equation}
where $q$ represents the fraction of loss mass during this region. Notice that $q$ enters linearly in~\eqref{eq:tauq} due to the absence of instantaneous mass tracking of the Hawking evaporation during the semiclassical phase~\cite{Dvali:2025ktz}. This is contrary to the common assumption made in the literature to derive constraints~\cite{Thoss:2024hsr,Chianese:2024rsn,Chianese:2025wrk,Boccia:2024nly,Boccia:2025hpm,Liu:2025vpz}.  For masses below $10^{10}\rm g$, constraints on $q$ arises from the fact that these BHs undergo mergers and therefore resume semiclassical evaporation in today's Universe, as pointed by~\cite{Zantedeschi:2024ram}. When considering this contribution, $t_{\rm formation}$ in Fig.~\ref{fig:qualitative} refers to the merger time and not to the cosmological formation time of the individual PBH.

\paragraph{The slow and the fast:} After the semiclassical emission ends, PBHs enter the memory burden phase, characterized by a suppressed emission rate, $\kappa < 1$. As discussed in previous sections, the transition to the memory burden floor emission, $\kappa \simeq S^{-k}$, can occur either slowly or rapidly. These two scenarios are illustrated in Fig.~\ref{fig:qualitative} by the solid cyan and blue lines, respectively. In the slow transition scenario, the key parameter depends on the duration of the phase. If the floor emission is reached within the age of the Universe, $t_0$, that is, when $\tau_{\rm SC}/(2t_0)\delta \simeq S^{-k}$, then $\delta$ is the only relevant parameter. However, if the slow emission phase is shorter, both $\delta$ and $k$ become important, as discussed in section~\ref{sec:theslow}. In the fast transition scenario, where the transition is effectively instantaneous ($\delta \to 0$), the only relevant parameter is $k$.

\smallskip
In the following sections, we systematically examine the theoretical and phenomenological signatures associated with the fast, slow, and merging scenarios, characterized by the parameters $k$, $\delta$, and $q$, respectively. We analyze the constraints on these parameters derived from a range of experiments, spanning the full spectrum of PBH masses within the memory-burden phase. In addition to the memory-burden parameters, we also consider the common parameter $f_{\rm PBH}$, which in some cases may exhibit degeneracy with the others. Unless otherwise stated, we assume that the parameter space under consideration corresponds to regions where the three scenarios remain non-degenerate.

\section{Phenomenological signatures}\label{sec:phenosign}

When a non-rotating, uncharged PBH with mass \( M_{\rm PBH} \) enters the burdened phase, the emission rate of a particle species \( i \) can be expressed as

\begin{equation}
    \frac{\dd^2N_{i}}{\dd E\,\dd t} = \xi\, \frac{\dd^2N^{\rm SC}_i}{\dd E\,\dd t} \ ,
    \label{eq:particleratetime}
\end{equation}
where \( {{\rm d}^2N^{\rm SC}_i}/{({\rm d}E \, {\rm d}t)} \) denotes the standard semi-classical Hawking emission, which peaks at the Hawking temperature \( T_H \):
\begin{equation}
    \frac{\dd^2N^{\rm SC}_i}{\dd E\,\dd t} = \frac{g_i}{2\pi} \frac{\mathcal{F}(M_{\rm PBH},x_i)}{e^{x_i M_{\rm PBH}/T_H} + 1} \ ,
\end{equation}
with \( g_i \) being the number of internal degrees of freedom of the species \( i \), and \( \mathcal{F}(M_{\rm PBH},x_i) \) the gray-body factor. Here $x_i$ is the energy fraction which depends on the memory burdened scenario. For the decay scenario, both the slow and fast, $x_i=E_i/M_{\rm PBH}$, while for the merger scenario $x_i=2E_i/M_{\rm PBH}$.
In~\eqref{eq:particleratetime}, the deviation from the semiclassical Hawking emission is encapsulated by the time-dependent function $\xi$, whose specific form depends on the scenario under consideration.
\begin{equation}
\begin{split}
    &\xi^{\rm {decay}} = 
   \max(S^{-k},\delta~\tau_{\rm SC}/2t)\quad \text{for } t\gtrsim \,\tau_{\rm SC}/2\ , \\
    &\xi^{\rm{merger}} = \frac{R_{\rm PBH}(f_{\rm PBH},t) ~q~\tau_{\rm SC}}{f_{\rm PBH}~\rho_{\rm DM,0}} M_{\rm {PBH}}  \ ,
    \end{split}
    \label{eq:cases}
\end{equation}
where $R_{\rm PBH}$ is defined in~\eqref{eq:mergerrate} and $\rho_{\rm DM,0}=\Omega_{\rm DM}\rho_{c,0}\simeq1.26\times 10^{-6}\,\si{GeV/cm^{3}}$~\cite{Planck:2018vyg} gives the today's DM energy density. In the semi-classical limit $\xi=1$. In this work, we analyze the decay scenario—both the \textit{fast} and \textit{slow} cases—and the \textit{merger} scenario separately. However, if decay and merger processes occur simultaneously, the total suppression factor is given by the sum $\xi^{\rm decay} + \xi^{\rm merger}$. To evaluate the initial particle emission rate in~\eqref{eq:particleratetime}, we use the numerical code {\tt BlackHawk 2.3}~\cite{Arbey:2019mbc, Arbey:2021mbl}, which accounts for the emission of all SM degrees of freedom. We obtain the total particle spectrum, including secondary emissions, by selecting the {\tt HDMSpectra} option~\cite{Bauer:2020jay} for PBH masses below $10^{10}\,$g, and the {\tt Pythia} option for larger masses. 
We point out that some confusion exists in the literature regarding the interpretation of the {\tt BlackHawk 2.3} output when using these two options. We clarify this in Appendix~\ref{app:AppBlackHawkclar}, and refer to~\cite{Bauer:2020jay} for a discussion of the main spectral differences between the two methods in the regime $E / \Lambda(M_{\rm PBH}) \lesssim 10^{-4}$.

\medskip
Given the spectrum at the production of a stable species $i$, the key observable relevant for cosmic-ray phenomenology is the differential flux $ \dd\Phi_i/\dd E_i$ observed at the Earth location. In this work, we assume a monochromatic mass distribution for the PBH population. Furthermore we
focus specifically on the emission of photons ($i=\gamma$) and neutrinos ($i=\nu$).

\smallskip
The prompt Galactic differential flux from a given angular region in the sky $\Delta\Omega = \int \cos b\, \dd b\, \dd \ell$, where $b$ and $\ell$ are the Galactic latitude and longitude coordinates respectively, is given by 
\begin{equation}
\frac{{\rm d}\Phi_{i,\rm gal}}{{\rm d}E_i \Delta\Omega} = \frac{1}{4\pi} \frac{f_{\rm PBH}}{M_{\rm PBH}} \, \frac{\dd^2N_i}{\dd E\,\dd t} \bar J \ , \qquad \mbox{with } \bar J = \frac1{\Delta\Omega}\int_{\Delta\Omega} \hspace{-0.2cm}\dd\Omega \int_0^\infty \hspace{-0.2cm}\dd s\, \rho_{\rm DM}(r(s, \psi))\, e^{-\tau_i(E_i,s)} \ .\label{eq:galFlux}
\end{equation}
Here, $\bar{J}$ denotes a generalized, averaged $J$-factor that integrates the intervening matter along the line of sight (parameterized by the variable $s$\footnote{The galactocentric distance in terms of $s$ is given by $r(s, \psi) = \sqrt{r_\odot^2 + s^2 - 2 r_\odot s \cos \psi}$, with $r_\odot = 8.3~\mathrm{kpc}$ and $\cos \psi = \cos b \cos \ell$.}) and over the angular region $\Delta\Omega$, while accounting for potential attenuation of the flux within the Galactic halo.  Throughout this work, we adopt a Navarro–Frenk–White (NFW) profile for the DM distribution~\cite{Navarro:1996gj}:
\begin{equation}
    \rho_{\rm DM}(r) = \frac{\rho_s}{(r/r_s)\left(1 + r/r_s\right)^2} \ ,
\end{equation}
where we fix the scale radius $r_s=25~\si{kpc}$ and we determine the scale density $\rho_s$ by requiring the local DM energy density to be $\rho_{\odot}=0.4\,\si{GeV/ cm^{3}}$~\cite{Riccardo_Catena_2010, Salucci_2010, Iocco_2015}. The optical depth $\tau_i$ in~\eqref{eq:galFlux} accounts for absorption along the line of sight. For $i = \nu$, the Milky Way is effectively transparent at all neutrino energies, such that $e^{-\tau_\nu} \equiv 1$. In contrast, for $i = \gamma$, the $\gamma$-ray flux is attenuated due to electron-positron pair  production (PP) on ambient photon fields in the Galaxy. In the energy range relevant to our analysis, the dominant photon background responsible for this absorption is the CMB, whose energy density significantly exceeds that of starlight and infrared fields. Consequently, substantial attenuation arises at photon energies around $E_\gamma \sim \mathcal{O}(10^6~\mathrm{GeV})$. This absorption is incorporated by introducing an exponential suppression factor, where the optical depth $\tau_\gamma\equiv \tau_{\gamma\gamma}^{\rm CMB}$ is analytically estimated  following the prescription in~\cite{Esmaili_2015}.

The secondary photon flux generated through Inverse Compton Scattering (ICS) of prompt high-energy electrons on the Galactic photon background is negligible. This is because the prompt emission primarily originates from hadronic cascades, resulting in a relatively soft spectrum that dominates over the ICS contribution.

\smallskip
Similarly to the case of decaying DM, we must also account for the contribution to the flux from evaporating PBHs throughout the history of the Universe. These emissions contribute to an isotropic component of the total observed flux intensity. As before, it is useful to distinguish between neutrino and photon emissions. 

The former propagate freely; therefore, the extragalactic contribution to the neutrino differential flux is given by:

\begin{figure*}[t!]
    \centering
        \begin{subfigure}[b]{0.49\textwidth}
        \centering
        \includegraphics[width=\linewidth]{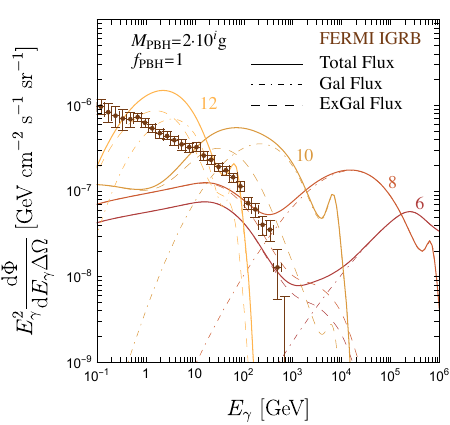}
    \end{subfigure}
    \,
    \begin{subfigure}[b]{0.49\textwidth}
        \centering
        \includegraphics[width=\linewidth]{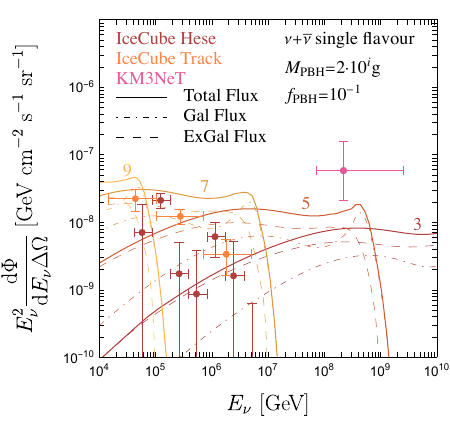}
    \end{subfigure}

\caption{Comparison of the full-sky differential fluxes of photons (left) and neutrinos (right) in the \textit{merger} scenario. Solid lines represent the total flux, while dashed and dot-dashed lines indicate the extragalactic and Galactic contributions, respectively. \textbf{Left panel:} Theoretical $\gamma$-ray flux computed for different values of $M_{\rm PBH}$, assuming $f_{\rm PBH} = 1$, compared with \textsc{Fermi}-data (brown points) for the IGRB. \textbf{Right panel:} Theoretical single-flavor neutrino flux for the same benchmark masses, assuming $f_{\rm PBH} = 10^{-1}$, compared with experimental data from \textsc{IceCube-HESE} (red points), \textsc{IceCube-Track} (orange points), and \textsc{KM3NeT} (pink point).
}

    \label{fig:FluxesMerger}
\end{figure*} 

\begin{equation}
   \frac{{\rm d} \Phi_{\nu,\rm{egal}}}{{\rm d}E_\nu} = \frac{f_{\rm PBH}}{M_{\rm PBH}}\,  \rho_{\rm{DM},0} \int_{0}^{z_f} \frac{{\rm d}z}{H(z)}\, \frac{{\rm d}^2N_\nu\left(E_\nu(1 + z)\right)}{{\rm d}E_\nu\, {\rm d}t} \, .
\label{eq:egalnu}
\end{equation}
Here, $H(z) = H_0 \sqrt{\Omega_M (1+z)^3 + \Omega_r (1+z)^4 + \Omega_\Lambda} $ is the Hubble expansion rate as a function of redshift $z$, where $H_0$ is its present-day value. The parameters $\Omega_{\rm DM}$, $\Omega_M$, $\Omega_r $, and $\Omega_\Lambda$ represent the present-day energy density fractions of DM, total matter, radiation, 
and the cosmological constant, respectively. The redshift integral extends up to the PBH formation redshift $z_f$. We checked that in all the scenarios the final constraints remain essentially unchanged as long as the upper limit of integration exceeds $z \sim 1000$.

In contrast, photons are subject to significant absorption processes. As a result, the extragalactic $\gamma$-ray emission consists of two distinct main contributions: the first is composed of primary $\gamma$-rays that survive attenuation during propagation, while the second arises from secondary emission produced by electromagnetic (EM) cascades initiated by the absorbed photons:

\begin{equation}
    \frac{{\rm d} \Phi_{\gamma, \mathrm{egal}}}{{\rm d} E_\gamma} = 
    \frac{{\rm d} \Phi_{\gamma, \mathrm{egal}}^{\rm att}}{{\rm d} E_\gamma} + 
    \frac{{\rm d} \Phi_{\gamma, \mathrm{egal}}^{\rm EM}}{{\rm d} E_\gamma} \,.
\end{equation}

The attenuated flux of photons emitted with energy \( E'_\gamma \) and observed today with redshifted energy \( E_\gamma = E'_\gamma / (1+z) \) is given by
\begin{equation}
    \frac{{\rm d} \Phi_{\gamma, \mathrm{egal}}^{\rm att}}{{\rm d} E_\gamma} = 
    \frac{f_{\rm PBH}}{M_{\rm PBH}}\, \rho_{\rm DM,0} 
    \int_0^{z_{\rm max}} \frac{{\rm d} z}{H(z)}\, 
    \frac{{\rm d}^2 N_\gamma\left(E_\gamma (1 + z)\right)}
         {{\rm d} E_\gamma\, {\rm d} t} \, 
    e^{-\tau_{\gamma}(E_\gamma,z)} \ ,
    \label{eq:egalgamma}
\end{equation}
where the exponential factor accounts for the absorption occurring within the energy range of interest. This arises from PP on the ambient photon background radiation (PBR), which is primarily composed of the CMB and the extragalactic Background Light (EBL). At high photon energies, $E_\gamma$ above tens of TeV, PP predominantly occurs through interactions with CMB photons, while at lower energies it is mainly driven by scattering on EBL photons. 
Additionally, after a PP event, the resulting electron-positron pairs can upscatter background photons via ICS, typically initiating an EM cascade. This process can repeat multiple times and leads to a significant enhancement of the $\gamma$-ray flux, especially at low energies. We simulate both the attenuation of primary high-energy photons and the resulting EM cascade using the \texttt{$\gamma$-CascadeV4} code~\cite{Blanco:2018bbf, Capanema:2024nwe} in the \textit{on-the-spot} approximation, adopting the best-fit EBL model from~\cite{Saldana-Lopez:2020qzx}. The integration is performed up to a redshift of \(z_{\rm max} = 10\), which adequately captures the bulk of the total emission. At higher redshifts, the Universe becomes increasingly opaque to photons with energies $E_\gamma \gtrsim 1~\si{TeV}$~\cite{Cirelli:2010xx, Slatyer:2009yq}. We stress that in the \texttt{$\gamma$-CascadeV4} code, the EM cascade is always initiated by PP on the EBL, which becomes efficient above \SI{10}{TeV}. However, an evaporating PBH emits all particle species democratically, including primary electrons and positrons. As a result, the EM cascade can also be triggered by these leptons. Consequently, we expect at most a correction of a factor of 2 to the predicted \(\dd \Phi^{\rm EM}_{\gamma,\rm{egal}}/\dd E_\gamma\).

\medskip
In conclusion, the total differential flux from a given angular region consists of both the Galactic and isotropic extragalactic contributions. It can be expressed as:
\begin{equation}\label{eq:tot_flux}
    \frac{{\rm d}\Phi_i}{{\rm d}E_i \Delta\Omega} \equiv \frac{{\rm d} \Phi_{i,\rm gal}}{{\rm d}E_i\Delta\Omega}+\frac1{4\pi} \frac{{\rm d} \Phi_{i,\rm egal}}{{\rm d}E_i} \ .
\end{equation}

As we will discuss in more detail in the following section, it is important to stress that different experiments observe different regions of the sky, and the signal-to-noise ratio generally depends on both the Galactic component—through the choice of the averaged J-factor $\bar{J}$ (analogous to the one of decaying DM)—and the isotropic extragalactic component. Unlike annihilating DM, the extragalactic emission is not substantially boosted from structure formation below $z\sim 30$. The only relevant component is the smooth cosmological one, whose redshift dependence varies depending on the scenario under study. One can move from the standard decay scenario, where the emissivity is simply proportional to the cosmological DM energy density \textit{i.e.}, it scales as $(1+z)^3$, to the merger scenario, where the PBH merger rate $R_{\rm PBH}$ introduces an additional clumpiness contribution, given by~\eqref{eq:mergerrate}. This contribution arises from binary formation processes that take place in the early Universe. We ignore extra boost factors from early three-body formation and all late-time channels which are generically poorly known. In summary, the cosmological component can contribute substantially to the total flux, and therefore the optimal observational strategy involves targeting a large portion of the sky.

In Fig.~\ref{fig:FluxesMerger}, we present an illustrative example (\textit{merger} scenario) in which we compare the full sky flux (solid lines), obtained by multiplying~\eqref{eq:tot_flux} by $\Delta \Omega = 4\pi$, with the corresponding experimental data. The dashed and dot-dashed lines represent the extragalactic and the Galactic contributions, respectively. In the left panel, the theoretical $\gamma$-ray flux, computed for different values of $M_{\rm PBH}$ and assuming $f_{\rm PBH} = 1$, is superimposed on the \textsc{Fermi} data coming from the measurement of the Isotropic $\gamma$-ray Background (IGRB)~\cite{Fermi-LAT:2014ryh}. As one can see, at low energies and for small PBH masses, the photon flux is dominated by the EM cascade component, which exhibits a universal peak slightly below 100~GeV, corresponding to the energy range where the cosmological flux is no longer significantly absorbed. This enhances the sensitivity of current observations to such small masses, as the resulting flux falls within the experimentally probed energy window. Similarly, in the right panel, we compare the theoretical neutrino flux (single flavor), calculated using the same benchmark masses and assuming $f_{\rm PBH} = 10^{-1}$, with \textsc{IceCube} data~\cite{Abbasi:2021qfz, IceCube:2020wum}. It is evident that the main contribution is the extragalactic component, since neutrinos do not undergo absorption during their propagation through the Universe.
The same key features are also present in the other scenarios; therefore, in the next section, we do not explicitly show the separate contributions to the total flux.

\begin{figure*}[t!]
    \centering
        \begin{subfigure}[b]{0.49\textwidth}
        \centering
        \includegraphics[width=\linewidth]{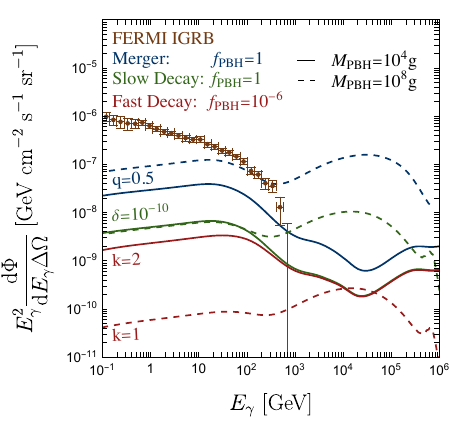}
    \end{subfigure}
    \,
    \begin{subfigure}[b]{0.49\textwidth}
        \centering
        \includegraphics[width=\linewidth]{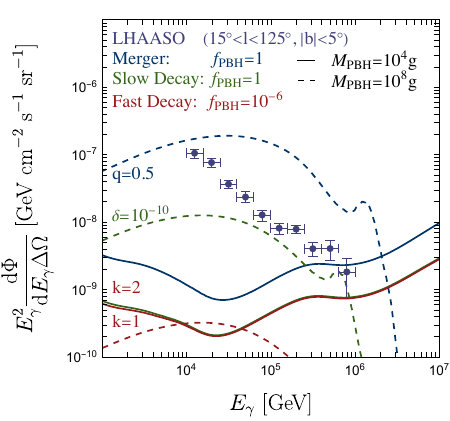}
    \end{subfigure}

    \vspace{0.2cm}

    \begin{subfigure}[b]{0.49\textwidth}
        \centering
        \includegraphics[width=\linewidth]{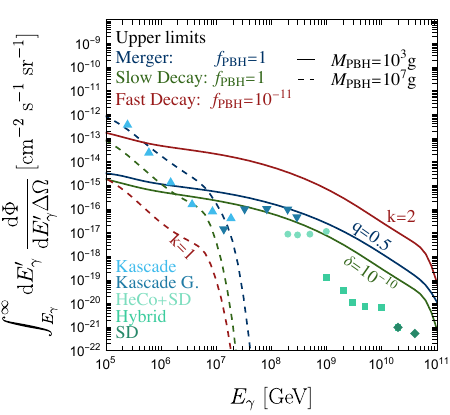}
    \end{subfigure}
    \,
    \begin{subfigure}[b]{0.49\textwidth}
        \centering
        \includegraphics[width=\linewidth]{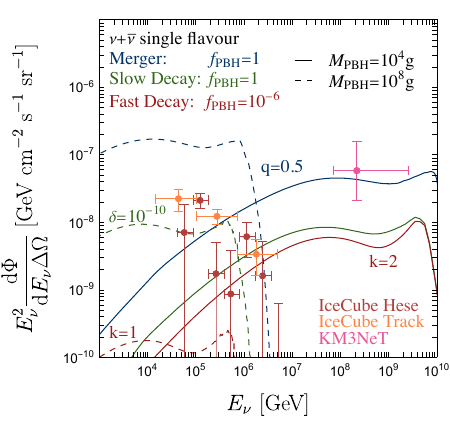}
    \end{subfigure}

    \caption{Comparison between the theoretical flux (Galactic + extragalactic components) and indirect detection data for different PBH masses and for the three scenarios considered: \textit{fast} (red), \textit{slow} (green), and \textit{merger} (blue). \textbf{Top-left panel:} Differential photon flux compared with \textsc{Fermi} data (brown points), assuming $f_{\rm PBH} = 1$ for the \textit{merger} and \textit{slow} scenarios, and $f_{\rm PBH} = 10^{-6}$ for the \textit{fast} scenario. Solid and dashed lines correspond to $M_{\rm PBH} = 10^4\,\si{g}$ and $10^8\,\si{g}$, respectively. \textbf{Top-right panel:} Same flux predictions compared with \textsc{LHAASO} data (dark purple points) in the inner Galactic region. \textbf{Bottom-left panel:} Integrated photon flux compared with upper limits from \textsc{Kascade} (light blue triangles), \textsc{Kascade-Grande} (blue triangles), and \textsc{Auger} (shades of aquamarine). Theoretical integrated fluxes are shown for $M_{\rm PBH} = 10^3\,\si{g}$ (solid lines) and $10^7\,\si{g}$ (dashed lines), assuming $f_{\rm PBH} = 1$ for the \textit{merger} and \textit{slow} scenarios, and $f_{\rm PBH} = 10^{-11}$ for the \textit{fast} scenario. \textbf{Bottom-right panel:} Theoretical single-flavor neutrino differential flux compared with \textsc{IceCube} and \textsc{KM3NeT} data, using the same parameters and PBH masses as in the top panels.
    }
    \label{fig:fluxes}
\end{figure*}

\section{Constraints}\label{sec:constr}
In this section, we examine several classes of constraints arising from both indirect detection and cosmological probes. We first use the measured diffuse emission of $\gamma$-rays (Sec.~\ref{subsubsec:gammarayconstr}) and neutrinos (Sec.~\ref{subsubsec:neutrinoconstr}) to place constraints on the parameters $k$, $\delta$, and $q$ in the range of PBH masses where the memory burden effect is relevant, i.e.~$10\,\si{g} \lesssim M_{\rm PBH} \lesssim 10^{15}\,\si{g}$. The lower bound of this mass range is set for practical reasons: in the energy window probed by current experiments, the theoretical prompt fluxes computed by {\tt BlackHawk} become unreliable for lighter PBHs. Furthermore, the inflationary production of PBHs generally requires $M_{\rm PBH} \gtrsim 1\,\si{g}$~\cite{Carr:2009jm}, making the study of lighter PBHs less motivated. In Sec.~\ref{subsec:cosmoprobes}, we also derive cosmological constraints on the free parameters of each scenario within the memory burden mass window and, where possible, compare them with existing semiclassical cosmological limits, which however extend up to $M_{\rm PBH} \sim 10^{17}\,\si{g}$.

\subsection{Indirect detection probes}
\label{subsec:IDprobes}
To derive the indirect detection bounds, we first outline the statistical analysis method, in which we treat separately the experiments that provide actual data and those that only set upper limits on the integrated fluxes. We then apply this method to the three relevant scenarios in order to constrain the free parameters of each as a function of the PBH mass.
\smallskip
For experiments with actual data $d$, we compare the total predicted flux with the observed data in each energy bin. This comparison is performed by introducing the following likelihood function~\cite{Arguelles:2019ouk}:

\begin{equation}\label{eq:likelihood}
    \mathcal{L}(\mu) \equiv 
\begin{cases}
    \prod_i \mathcal{P}_i(d|\mu) & \text{for } \mu >d \\
    1 & \text{for } \mu \le d\, 
\end{cases} \ ,
\end{equation}
Here, $\mu$ represents the theoretically predicted differential flux, and the product runs over the number of bins that satisfy the condition $\mu > d$. Depending on the scenario under consideration, $\mu$ depends on a specific number of free parameters: two independent parameters for the fast decay scenario, $\mu(f_{\rm PBH}, k; M_{\rm PBH})$; one for the slow case, $\mu(f_{\rm PBH} \, \delta; M_{\rm PBH})$; and two for the merger scenario, $\mu(f_{\rm PBH}, q; M_{\rm PBH})$. The function $\mathcal{P}_i$ denotes the probability distribution, which we assume to be Gaussian for the datasets considered.  To extract the constraints, we construct the test statistic $\Delta\chi^2 \equiv -2\ln\left({\mathcal{L}/\mathcal{L}_0}\right)$ for each PBH mass and apply Wilks' theorem. Here, $\mathcal{L}_0$ represents the likelihood of the null hypothesis, which depends on the statistical analysis.
 In the following, we always assume one degree of freedom. This condition is automatically satisfied for the slow scenario, while for the fast and merging scenarios, we fix one parameter at a time, as explained in more detail below. Under this assumption, we impose $\Delta\chi^2 = 2.71$ to determine the 95\% confidence level interval for the relevant free parameter in each scenario.

\smallskip
For experiments that only provide upper limits, such as most of the ultrahigh-energy (UHE) photon experiments, the key theoretical quantity is the integrated photon flux, computed from~\eqref{eq:tot_flux} by choosing $i = \gamma$, and defined as
\begin{equation}\label{eq:int_flux}
\Phi_{\rm int}(E_{\gamma}) = \int_{E_{\gamma}}^{+\infty} \frac{{\rm d}\Phi_{\gamma}}{{\rm d}E^\prime_{\gamma} \, \Delta\Omega} \, {\rm d}E^\prime_{\gamma} \ .
\end{equation}
Depending on the scenario under consideration, we compare~\eqref{eq:int_flux}, bin by bin, with the upper limits provided by each collaboration, assuming, as explained above, one parameter at a time.

\smallskip
We now outline the relevant experimental measurements used to derive constraints on the photon and neutrino fluxes.
 
\subsubsection{Constraints from $\gamma$-ray experiments}
\label{subsubsec:gammarayconstr}

\paragraph{Fermi-LAT:} We use the measurements of the IGRB spectrum in the energy range $0.1\,\mathrm{GeV} \leq E_\gamma \leq 820\,\mathrm{GeV}$, as reported in Ref.~\cite{Fermi-LAT:2014ryh}.  Given the isotropic nature of these measurements, we compute the total all-sky averaged $J$-factor. For energies well below the attenuation threshold, this yields $\bar{J}(E_\gamma \ll 10^6\,\mathrm{GeV}) \approx 2.25 \times 10^{22}\,\mathrm{GeV}/\mathrm{cm}^2$. The data are, for example, shown as brown dots in the top-left panel of Fig.~\ref{fig:fluxes}. We compare them with the theoretical fluxes predicted in various scenarios, considering two values of the PBH mass ($10^4\,\mathrm{g}$, solid line; $10^8\,\mathrm{g}$, dashed line). More specifically, the fluxes in the \textit{fast} scenario (red) are computed using $f_{\rm PBH} = 10^{-6}$ for both PBH masses, with $k = 2$ for $10^4\,\mathrm{g}$ and $k = 1$ for $10^8\,\mathrm{g}$. In the \textit{slow} scenario (green), we assume $f_{\rm PBH} = 1$ and $\delta = 10^{-10}$, while in the \textit{merger} scenario (dark blue), we adopt $q = 0.5$ and $f_{\rm PBH} = 1$.
    
\begin{figure*}[t!]
    \centering
    \begin{subfigure}[b]{0.5\textwidth}
        \centering
        \includegraphics[width=\linewidth]{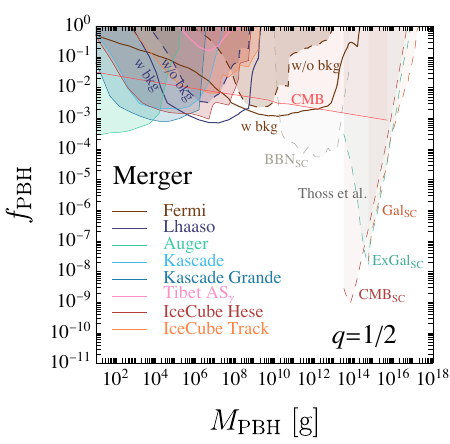}
    \end{subfigure}
    \hfill
    \begin{subfigure}[b]{0.48\textwidth}
        \centering
        \includegraphics[width=\linewidth]{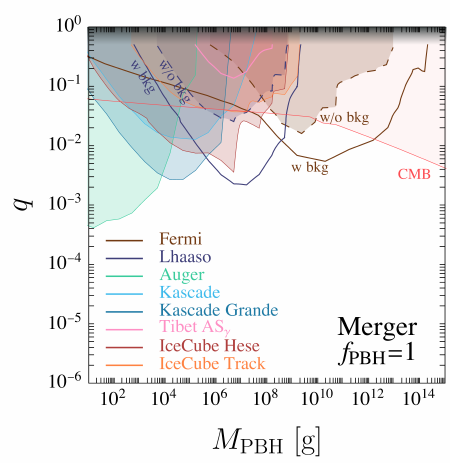}
    \end{subfigure}
    
    \caption{\textbf{Left panel}: Sensitivities of $ f_{\rm PBH} $ as a function of $M_{\rm PBH}$ in the \textit{merger scenario}, assuming a fixed relative evaporated mass fraction $q = 1/2$. Colored curves denote the sensitivity from different experiments: \textsc{Fermi} (brown), \textsc{Lhaaso} (dark purple), \textsc{Auger} (aquamarine), \textsc{Kascade} (light blue), \textsc{Kascade-Grande} (blue), \textsc{Tibet AS$\gamma$} (pink), and \textsc{IceCube} (HESE and Track datasets in dark red and orange, respectively). The semiclassical CMB, BBN, Galactic and extragalactic constraints are denoted by dashed dark red solid and black lines and were taken from~\cite{Thoss:2024hsr} (they are denoted by the pedix SC). The CMB bound from the \textit{merger} phase derived in this work is represented by the solid red line. \textbf{Right panel}: Same as the left panel, but assuming $f_{\rm PBH} = 1$ and deriving constraints on the evaporated mass fraction $q$. The dark shadow region indicates the limit $q\gtrsim1/2$ c.f.,~\eqref{qdef}. We stress that, as discussed in Section~\ref{sec:essence}, these constraints are computed assuming non-rotating BHs.
    } 
    \label{fig:money_plot_merger_all}
\end{figure*}

\begin{figure*}[t!]
    \centering
    \begin{subfigure}[b]{0.51\textwidth}
        \centering
        \includegraphics[width=\linewidth]{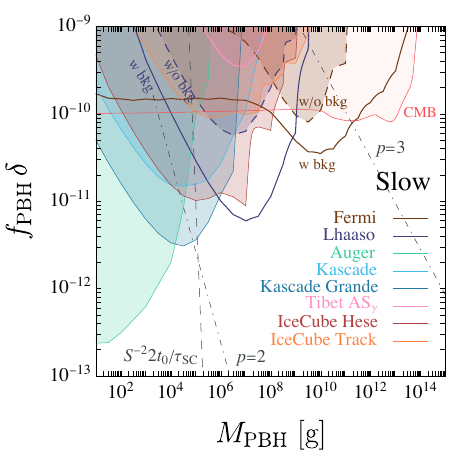}
    \end{subfigure}
    \hfill
    \begin{subfigure}[b]{0.475\textwidth}
        \centering
        \includegraphics[width=\linewidth]{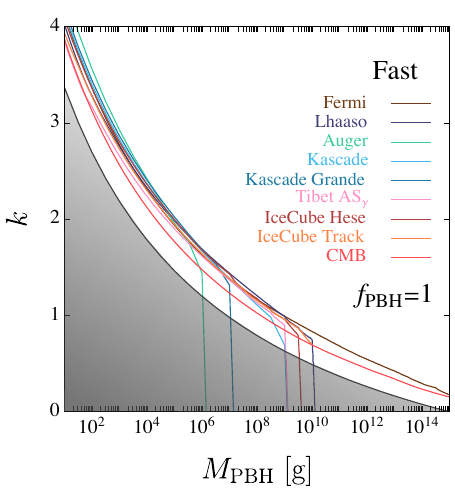}
    \end{subfigure}
    
    \caption{
    The color code for the experimental sensitivities matches that of Fig.~\ref{fig:money_plot_merger_all}. \textbf{Left panel:} Sensitivity on $f_{\rm PBH} \delta$ as a function of the PBH mass in the \textit{slow} decay scenario. The red solid line represents the CMB constraint derived in Subsection~\ref{subsec:cosmoprobes}. Dashed line corresponds to $\delta=S^{-2}2t_0/\tau_{\rm SC}$ and indicates the mass for which PBHs are reaching the memory burden ``floor'' today at $k=2$. The dot dashed lines correspond to the $p=2$ and $p=3$ benchmarks. \textbf{Right panel:} Exclusion regions in the $(M_{\rm PBH}, k)$ parameter space for the \textit{fast} decay scenario, assuming $f_{\rm PBH} = 1$. The gray shaded region indicates the part of the parameter space for which $\tau \lesssim t_0$ c.f., \eqref{tMB}.}
    
    \label{fig:money_plot_decay_all}
\end{figure*}

    We perform two distinct analyses: $i$) a \textit{background-agnostic} analysis, in which we use the likelihood from~\eqref{eq:likelihood}, assuming that $\mu$ corresponds solely to the all-sky DM differential flux, as given in~\eqref{eq:tot_flux}. The null hypothesis likelihood in this case is $\mathcal{L}_0 \equiv \mathcal{L}(\mu = d)$; $ii$) a \textit{background-inclusive} analysis, in which we include the background, modeled as a power law with an exponential cut-off, as described in Ref.~\cite{Fermi-LAT:2014ryh}. In this case, the likelihood in~\eqref{eq:likelihood} is evaluated with $\mu$ that includes both the DM-induced flux and the background contribution, while the null hypothesis likelihood accounts only for the background. By construction, the second method yields more stringent constraints and is more physically motivated, as the DM-induced flux alone does not provide a satisfactory fit to the data. The resulting bounds at 95\% confidence level are shown in brown in Fig.~\ref{fig:money_plot_merger_all} for the merger scenario and in Fig.~\ref{fig:money_plot_decay_all} for both the \textit{fast} and \textit{slow} decay scenarios. The dashed lines refer to the background-agnostic analysis, while the solid lines represent the results from the background-inclusive analysis.

    More specifically, the left panel of Fig.~\ref{fig:money_plot_merger_all} shows the results in the $(M_{\rm PBH}, f_{\rm PBH})$ plane, assuming an evaporated mass fraction $q = 0.5$. As one can see, the IGRB measured by \textsc{Fermi} provides very stringent constraints in the high-mass regime, extending up to $M_{\rm PBH} \sim 10^{14}~\mathrm{g}$. In the mass window $(10^9\text{--}10^{10})~\mathrm{g}$, the background-inclusive analysis yields the strongest limit to date, reaching $f_{\rm PBH} \simeq 10^{-3}$. 
    In the right panel of the same figure, we show the constraints in the $(M_{\rm PBH}, q)$ plane, assuming a PBH fraction $f_{\rm PBH} = 1$. The background-inclusive analysis of the IGRB measured by \textsc{Fermi} alone is capable of completely ruling out the case $q = 0.5$ and $f_{\rm PBH} = 1$ across the entire PBH mass range where the memory burden effect is relevant. As before, it provides the strongest constraint in the mass window $(10^9\text{--}10^{10})~\mathrm{g}$, reaching sensitivity to the mass fraction $q\simeq5 \times 10^{-3}$. In both cases, the constraining power in the low-mass tail (below $10^{8}~\mathrm{g}$, corresponding to a prompt peak emission at around $10^6~\mathrm{GeV}$) is largely controlled by the extragalactic secondary EM cascade, which exhibits a universal peak slightly below 100~GeV and lies well within the \textsc{Fermi} energy window. As shown in the top left panel of Fig.~\ref{fig:fluxes}, the normalization of the extragalactic contribution depends on $M_{\rm PBH}$, which explains why the low-energy tail of the bound does not form a plateau.

   The left panel of Fig.~\ref{fig:money_plot_decay_all} focuses on the \textit{slow} decay case, where the results are displayed in the $(M_{\rm PBH}, f_{\rm PBH}\delta)$ plane. In this scenario, the background-inclusive analysis of the IGRB provides the strongest limit to date across a broad range of high PBH masses, above $10^9~\mathrm{g}$. The best sensitivity reaches $f_{\rm PBH}\delta \simeq 4\times 10^{-11}$ at a PBH mass around $10^{10}~\mathrm{g}$. In this region, the \textsc{Fermi} bound surpasses those from other $\gamma$-ray observations (particularly \textsc{Lhaaso}), as well as limits from the CMB. As in the merger scenario, the constraining power below $10^{8}~\mathrm{g}$ is controlled by the extragalactic secondary EM cascade. This time, as shown in the top-left panel of Fig.~\ref{fig:fluxes}, it is the full extragalactic secondary EM cascade that is universal, giving rise to a plateau in the low-mass tail. Finally, the right panel of Fig.~\ref{fig:money_plot_decay_all} focuses on the \textit{fast} decay case, where the results (coming from the background inclusive analysis) are displayed in the $(M_{\rm PBH}, k)$ plane. In this case, for $M_{\rm PBH} \gtrsim 10^9\,\mathrm{g}$, \textsc{Fermi} is the leading experiment; however, it constrains $k\lesssim 1$, which is not particularly compelling from a theoretical perspective. On the other hand, the theoretically motivated benchmark $k = 2$ is reached for $M_{\rm PBH} \simeq 6\times 10^4\,\mathrm{g}$, providing stronger constraints than cosmological probes, although still weaker than other indirect detection experiments.

    In general, as shown in both Fig.~\ref{fig:money_plot_merger_all} and Fig.~\ref{fig:money_plot_decay_all}, the \textit{background-agnostic} analysis yields weaker bounds. At the point of maximum sensitivity, the two bounds are of the same order, as only a few low precise high-energy data points dominate the constraining power. The main difference emerges at the edges of the PBH mass window, where the $\gamma$-ray spectrum falls within the \textsc{Fermi} energy range. In these regions, more data points contribute to the constraint, and as a result, the \textit{background-inclusive} analysis significantly strengthens the bound thanks to the high precision of \textsc{Fermi} data points. As previously noted, at low PBH masses the EM cascade lies within the \textsc{Fermi} window, while at higher masses, is the Galactic contribution to be well inside the \textsc{Fermi} energy window.

\paragraph{LHAASO:} We use the measurements of the diffuse $\gamma$-ray emission from the Galactic plane reported by the KM2A site of the \textsc{Lhaaso} experiment~\cite{LHAASO:2023gne}. These observations, covering the energy range from 0.1~PeV to 1~PeV, focus on two regions of the Galactic plane: an inner region defined by the angular aperture $-5^\circ \le b \le 5^\circ$, $15^\circ \le \ell \le 125^\circ$, and an outer region defined by $-5^\circ \le b \le 5^\circ$, $125^\circ \le \ell \le 235^\circ$. The resulting $J$-factors in~\eqref{eq:galFlux} are $\bar{J}_{\rm in}(E_{\gamma}\ll10^6\,\text{GeV})\approx 2.72 \times 10^{22}\,\mathrm{GeV}/\mathrm{cm}^2$ and $\bar{J}_{\rm out}(E_{\gamma}\ll10^6\,\text{GeV})\approx 1.22 \times 10^{22}\,\mathrm{GeV}/\mathrm{cm}^2$, for the inner and outer regions respectively. The \textit{inner Galactic region} dataset is shown as dark purple dots in the top-right panel of Fig.~\ref{fig:fluxes}, and is compared with theoretical flux predictions for the three scenarios under consideration. The label scheme and parameter choices are the same as those previously discussed for \textsc{Fermi}.

We apply the same analysis procedure as used for the \textsc{Fermi} dataset, performing both a \textit{background-agnostic} and a \textit{background-inclusive} analysis. In the latter case, we model the background as a simple power law, using the best-fit parameters provided in Ref.~\cite{LHAASO:2023gne}. The resulting 95\% confidence level sensitivities are shown in Figs.~\ref{fig:money_plot_merger_all} and~\ref{fig:money_plot_decay_all}, with the dark purple dashed and solid lines representing the background-agnostic and background-inclusive analyses, respectively. To this respect, we report the bounds obtained from the inner Galactic region, which provides slightly stronger limits with respect to the outer region.

For the \textit{merger} scenario, as shown in Fig.~\ref{fig:money_plot_merger_all}, the \textit{background-inclusive} analysis covers a wide mass range and provides the most stringent limits to date in the $(10^5\text{--}10^8)~\mathrm{g}$ range, surpassing the \textsc{Fermi} constraint. The best sensitivity is achieved at $M_{\rm PBH} \simeq 10^7\,\si{g}$, where the exclusion in the $(M_{\rm PBH}, f_{\rm PBH})$ plane reaches $f_{\rm PBH} \simeq 7 \times 10^{-4}$ (assuming $q = 1/2$), and in the $(M_{\rm PBH}, q)$ plane probes down to $q \simeq 2 \times 10^{-3}$ (assuming $f_{\rm PBH} = 1$).

In Fig.~\ref{fig:money_plot_decay_all} we focus on both the \textit{fast} (right panel) and the \textit{slow} (left panel) scenarios. In particular, for the \textit{slow} case (left panel), we see that the constraint resulting from the background-inclusive analysis covers a mass window from $10^3~\mathrm{g}$ to $10^9\,\mathrm{g}$. In the range between $10^6\,\mathrm{g}$ and $10^9\,\mathrm{g}$ \textsc{Lhaaso} is the leading experiment achieving its best sensitivity of $f_{\rm PBH}\delta \simeq 6\times 10^{-12}$ at $M_{\rm PBH} \simeq 10^7\,\mathrm{g}$. Unlike \textsc{FERMI}, here the low-mass tail is not affected by the characteristic plateau typically produced by the EM cascade. This difference arises because \textsc{FERMI} collects data over the full sky, making the isotropic extragalactic contribution a significant component of the total photon flux. In contrast, \textsc{Lhaaso} observes a much narrower region of the sky, $\Delta\Omega \ll 4\pi$, specifically aligned with the Galactic disk. As a result, the local Galactic contribution, accounted for by the $\bar{J}$ factor, plays a more prominent role in the experiment’s constraining power. In the right panel of Fig.~\ref{fig:money_plot_decay_all}, we present the sensitivity for the \textit{fast} decay scenario in the $(M_{\rm PBH}, k)$ plane assuming $f_{\rm PBH}=1$. As for the other scenarios, the constraining power extends up to $M_{\rm PBH}\simeq 10^{10}\,\mathrm{g}$, where the noticeable drop in sensitivity appears because no low-energy data are available. Furthermore, \textsc{Lhaaso} emerges as the leading experiment in the PBH mass range $(10^7$–$10^8)\,\mathrm{g}$, where $k\lesssim1$. The motivated benchmark value $k = 2$ is reached at $M_{\rm PBH} \simeq 8\times 10^4\,\mathrm{g}$.

As a general comment, we remark that including the background in the \textsc{Lhaaso} analysis leads to an improvement of approximately one order of magnitude at the minimum of the sensitivity curve, in contrast to the \textsc{Fermi} case. This difference arises because, as clearly shown in Fig.~\ref{fig:fluxes}, the \textsc{Fermi} reach is dominated by a few high-energy data points with relatively large statistical uncertainties. In contrast, the \textsc{Lhaaso} constraints are driven by a larger number of data points, which collectively enhance the overall sensitivity. At the edges of the constraint, the two analyses begin to provide a reach of the same order. This is because, at high PBH masses, only the first bins dominate the constraining power, while at low PBH masses, the energy spectrum is shifted outside the \textsc{Lhaaso} sensitivity window, and the resulting EM cascade is too soft and subdominant to yield a significant bound.

\paragraph{Tibet AS$_\gamma$:} We use measurements of diffuse  $\gamma$-rays from the Galactic Disk in the energy range $140~\text{TeV} < E_{\gamma} < 1.3~\text{PeV}$~\cite{TibetASgamma:2021tpz}, focusing on two regions of interest: an inner region defined by the angular range $-5^\circ \le b \le 5^\circ$, $25^\circ \le \ell \le 100^\circ$, and an outer region defined by $-5^\circ \le b \le 5^\circ$, $50^\circ \le \ell \le 200^\circ$. Focusing on these two spatial windows gives $\bar{J}_{\rm in}(E_{\gamma}\ll10^6\,\text{GeV})\approx 2.73 \times 10^{22}\,\mathrm{GeV}/\mathrm{cm}^2$ and $\bar{J}_{\rm out}(E_{\gamma}\ll10^6\,\text{GeV})\approx 1.57 \times 10^{22}\,\mathrm{GeV}/\mathrm{cm}^2$ for the $J$-factors of the inner ($\Delta\Omega_{\rm in}\approx 0.23\,\text{sr}$) and outer  ($\Delta\Omega_{\rm out}\approx 0.46\,\text{sr}$) regions, respectively. 

We perform only a \textit{background-agnostic} analysis using~\eqref{eq:likelihood}. For \textsc{Tibet-AS$_\gamma$}, the available data consist of only a few points with large statistical uncertainties; therefore, a \textit{background-inclusive} analysis is not justified, and any improvement would be marginal. Both the inner and outer regions yield comparable constraints and in Figs.~\ref{fig:money_plot_merger_all},~\ref{fig:money_plot_decay_all} we show with a pink solid line the resulting 95\% confidence level sensitivity corresponding to the inner region. For the \textit{merger} scenario (Fig.~\ref{fig:money_plot_merger_all}), the constraint covers the PBH mass window from $10^5\,\mathrm{g}$ to $10^8\,\mathrm{g}$, reaching its maximum sensitivity at $M_{\rm PBH} \simeq 4 \times 10^{-6}\,\mathrm{g}$, where the exclusion in the $(M_{\rm PBH},f_{\rm PBH})$ plane reaches $f_{\rm PBH} \simeq 0.2$ assuming $q = 1/2$, and in the $(M_{\rm PBH},q)$ plane probes down to $q \simeq 0.1$ assuming $f_{\rm PBH} = 1$. Fig.~\ref{fig:money_plot_decay_all} shows the results for the decay scenario. In the \textit{slow} decay case (left panel), the constraint covers a similar PBH mass range, with the strongest sensitivity at $M_{\rm PBH} \simeq 4 \times 10^{-6}\,\mathrm{g}$ and $f_{\rm PBH} \delta \simeq 3 \times 10^{-10}$. For the \textit{fast} decay scenario (right panel), the benchmark value $k = 2$ is reached at $M_{\rm PBH} \simeq 4 \times 10^4\,\mathrm{g}$.

\paragraph{KASCADE \& KASCADE-Grande:} We use the 90\% confidence level upper limits on the isotropic\footnote{Bounds from Extensive Air Shower detectors on the $\gamma$-ray fraction in the cosmic-ray flux are derived under the assumption of an isotropic flux. However, as pointed out in~\cite{Esmaili_2015}, this approximation becomes untenable in the energy range of interest due to the direction-dependent optical depth of the Galactic sky.} diffuse $\gamma$-ray flux in the energy range 100~TeV to 1~EeV, as reported by the \textsc{Kascade} and \textsc{Kascade-Grande} experiments~\cite{KASCADEGrande:2017vwf}. Following the procedure described earlier, we compute the integral flux from~\eqref{eq:int_flux} using the all-sky differential photon flux, and compare it with the experimental upper limits bin by bin. The data are shown as light blue (\textsc{Kascade}) and blue (\textsc{Kascade-Grande}) triangles in the bottom-left panel of Fig.~\ref{fig:fluxes}. They are compared with theoretical integrated flux predictions for the fast (red), slow (green), and merger (blue) scenarios. Solid and dashed lines correspond to two different PBH mass values, $10^3\,\mathrm{g}$ and $10^7\,\mathrm{g}$, respectively. For both the merger and slow scenarios, we set $f_{\rm PBH} = 1$ and choose $q = 0.5$ and $\delta = 10^{-10}$, respectively. In the fast decay scenario, we fix $f_{\rm PBH}=10^{-11}$, $k = 1$ for $M_{\rm PBH} = 10^7\,\mathrm{g}$ and $k = 2$ for $M_{\rm PBH} = 10^3\,\mathrm{g}$. 

The resulting upper limits are shown in Figs.~\ref{fig:money_plot_merger_all},~\ref{fig:money_plot_decay_all} with a light blue and blue line for \textsc{Kascade} and \textsc{Kascade-Grande}, respectively. More specifically, for the \textit{merger} scenario, in Fig.~\ref{fig:money_plot_merger_all}, the reach covers the PBH mass window from $10\,\mathrm{g}$ up to $10^6\,\mathrm{g}$ for \textsc{Kascade-Grande} and up to $10^7\,\mathrm{g}$ for \textsc{Kascade}. In the interval $(10^4-10^5)\,\mathrm{g}$ \textsc{Kascade-Grande} is the leading experiment with the best upper limit reached at $M_{\rm PBH}\simeq 2\times 10^4\,\mathrm{g}$, where the exclusion in the  $(M_{\rm PBH}, f_{\rm PBH})$ plane constrains $f_{\rm PBH}\simeq 7\times 10^{-4}$ assuming $q=1/2$, and in the  $(M_{\rm PBH}, q)$ plane probes down to $q\simeq 2\times 10^{-3}$ assuming $f_{\rm PBH}=1$. Coming to the \textit{slow} and \textit{fast} decay scenarios in Fig.~\ref{fig:money_plot_decay_all}, \textsc{Kascade} and \textsc{Kascade-Grande} cover the PBH mass window up to $10^7\,\mathrm{g}$ and $10^8\,\mathrm{g}$, respectively. In the \textit{slow} case (left panel), \textsc{Kascade-Grande} is the leading experiment in the $(10^4\text{--}10^6)\,\mathrm{g}$ mass window with the best sensitivity at $M_{\rm PBH}\simeq 3\times 10^4\,\mathrm{g}$ with $f_{\rm PBH}\delta\simeq 3\times 10^{-12}$. In the \textit{fast} decay case (right panel), the bound extends up to $10^{9}\,\mathrm{g}$ where the typical sensitivity drop occurs. The benchmark point $k=2$ is reached at $M_{\rm PBH}\simeq 10^{5}\,\mathrm{g}$ from both \textsc{Kascade} and \textsc{Kascade-Grande}.

\paragraph{Pierre Auger Observatory:} We use three independent datasets that provide 95\% confidence level upper limits on the UHE photon flux. These include: the HeCo+SD dataset~\cite{PierreAuger:2022uwd}, covering the energy range $10^{17}\,\text{eV} \lesssim E_{\gamma} \lesssim 10^{18}\,\text{eV}$; the Hybrid dataset~\cite{Savina:2021cva}, spanning $10^{18}\,\text{eV} \lesssim E_{\gamma} \lesssim 10^{19}\,\text{eV}$; and the Surface Detector (SD) data~\cite{PierreAuger:2022aty}, which focuses on $E_{\gamma} > 10^{19}\,\text{eV}$. The three datasets are represented by different shades of aquamarine, using points, squares, and rhombuses, as shown in the bottom-left panel of Fig.~\ref{fig:fluxes}. They are compared with theoretical integrated flux predictions for the three scenarios under consideration. The color scheme and parameter choices are the same as those used for \textsc{Kascade} and \textsc{Kascade-Grande}.

The combined upper limit derived from these datasets is shown in Figs.~\ref{fig:money_plot_merger_all},~\ref{fig:money_plot_decay_all} using an aquamarine line. For the \textit{fast}, the \textit{slow} and the \textit{merger}, \textsc{Auger} is the leading experiment in the low PBH mass window, $(10-10^4)\,\mathrm{g}$. For the \textit{merger} case, the maximal sensitivity in the $(M_{\rm PBH},f_{\rm PBH})$ plane reaches $f_{\rm PBH}\simeq 3\times 10^{-4}$ fixing $q=1/2$ and in the  $(M_{\rm PBH}, q)$ plane probes down to $q\simeq 4 \times 10^{-4}$ for $f_{\rm PBH}=1$. In Fig.~\ref{fig:money_plot_decay_all} we show the results for both the \textit{slow} case (left panel) and \textit{fast} decay case (right panel). In both the cases, the constraint extends up to $10^{5}\,\mathrm{g}$ and set the most stringent constraint to date in the mass window $(10-10^4)\,\mathrm{g}$. For the \textit{slow} case the maximal reach is $f_{\rm PBH}\delta\simeq 2\times 10^{-13}$, while for the \textit{fast} decay the motivated benchmark $k=2$ is reached at $M_{\rm PBH}\simeq 8\times 10^4\,\mathrm{g}$.

\subsubsection{Constraints from neutrino experiments}
\label{subsubsec:neutrinoconstr}

For the neutrino experimental data, we assume that the measured flux is approximately isotropic and does not trace the Galactic plane. Due to the complexity of background modeling, and the limited number and large statistical uncertainties of the available data points, we consistently adopt a \textit{background-agnostic} analysis based on Eq.~\eqref{eq:likelihood}, where $\mu$ denotes the isotropic all-sky differential muon-neutrino flux produced by PBHs.

\paragraph{IceCube 9.5 Year Dataset:} We make use of the astrophysical muon-neutrino flux measurement from the 9.5-year \textsc{IceCube} dataset~\cite{Abbasi:2021qfz}, derived from a high-purity sample of muon tracks produced by neutrinos in the 15 TeV $-$ 5 PeV energy range. In this case, the value of the $J$-factor is the same as the one quoted in the \textsc{Fermi} case, $\bar{J} \approx 2.25 \times 10^{22}\,\mathrm{GeV}/\mathrm{cm}^2$, with the important difference that it has no energy dependence, since neutrinos propagate unattenuated. The \textsc{IceCube} track dataset is shown as orange points in the bottom-right panel of Fig.~\ref{fig:fluxes}. These data are compared with theoretical flux predictions for the three scenarios under consideration. The color scheme and parameter choices are the same as those used for both \textsc{Fermi} and \textsc{Lhaaso}.

The resulting 95\% confidence level sensitivity is shown in Figs.~\ref{fig:money_plot_merger_all} and~\ref{fig:money_plot_decay_all} using an orange line. For the \textit{merger} scenario (Fig.~\ref{fig:money_plot_merger_all}), the constraints span the PBH mass range $(10^3$--$10^9)\,\mathrm{g}$, with maximal sensitivity reached around $M_{\rm PBH} \simeq 10^7\,\mathrm{g}$. In the $(M_{\rm PBH}, f_{\rm PBH})$ plane, the strongest bound corresponds to $f_{\rm PBH} \simeq 7 \times 10^{-3}$ assuming $q = 1/2$, while in the $(M_{\rm PBH}, q)$ plane the constraint reaches $q \simeq 3 \times 10^{-2}$ for $f_{\rm PBH} = 1$. In Fig.~\ref{fig:money_plot_decay_all}, we present the results for both the \textit{slow} (left panel) and \textit{fast} (right panel) decay scenarios. In the slow decay case, the maximal sensitivity is $f_{\rm PBH}\delta \simeq 9 \times 10^{-11}$, whereas in the fast decay case, the benchmark value $k = 2$ is reached at $M_{\rm PBH} \simeq 5 \times 10^4\,\mathrm{g}$.

    \paragraph{IceCube 7.5 Year Dataset:} We consider the measurement of the astrophysical neutrino flux from the 7.5-year High-Energy Starting Event (HESE) sample, for events with reconstructed energies above 60~TeV~\cite{IceCube:2020wum}. As for the \textsc{IceCube} track dataset, the HESE dataset is shown as red points in the bottom-right panel of Fig.~\ref{fig:fluxes}, following the same color scheme and parameter choices.
    
    The resulting 95\% confidence level sensitivity is shown in Figs.~\ref{fig:money_plot_merger_all} and~\ref{fig:money_plot_decay_all} using a dark red line. For the \textit{merger} scenario (Fig.~\ref{fig:money_plot_merger_all}), the constraints span the PBH mass range  up to $10^9\,\mathrm{g}$, with maximal sensitivity reached around $M_{\rm PBH} \simeq 10^7\,\si{g}$. In the $(M_{\rm PBH}, f_{\rm PBH})$ plane, the maximal reach corresponds to $f_{\rm PBH} \simeq  10^{-3}$ assuming $q = 1/2$, while in the $(M_{\rm PBH}, q)$ plane the constraint reaches $q \simeq 4 \times 10^{-3}$ for $f_{\rm PBH} = 1$. In Fig.~\ref{fig:money_plot_decay_all}, we present the results for both the \textit{slow} (left panel) and \textit{fast} (right panel) decay scenarios. In the slow decay case, the maximal sensitivity is $f_{\rm PBH}\delta \simeq 10^{-11}$, whereas in the fast decay case, the benchmark value $k = 2$ is reached at $M_{\rm PBH} \simeq  10^5\,\mathrm{g}$.
    
    We remark that the maximal sensitivity is achieved for $M_{\rm PBH} \simeq 10^7\,\mathrm{g}$, where the prompt neutrino flux peaks at $E_{\nu} \simeq 10^7~\si{GeV}$, corresponding to the lowest-energy point of the \textsc{HESE} dataset. A secondary sensitivity peak occurs at $M_{\rm PBH} \simeq 10^6\,\mathrm{g}$, where the prompt neutrino flux, peaking at $E_{\nu} \simeq 10^6~\si{GeV}$, is most constrained by the second-lowest data point. In general, for low PBH masses, the constraining power is primarily driven by the extragalactic contribution to the neutrino flux which, unlike the $\gamma$-ray flux, is not subject to absorption. Moreover, in both the \textit{merger} and \textit{slow} decay scenarios, the low-energy tails of the flux are enhanced due to the steeper redshift evolution of the source population.

\paragraph{KM3NeT PeV Event:} The \textsc{KM3NeT} collaboration recently reported the observation of a $\sim 220$\,PeV muon event (KM3--230213A), likely induced by a neutrino with energy $E_\nu \sim 110{-}790$\,PeV~\cite{KM3NeT:2025npi} with a differential flux $E^2{\rm d}\Phi/{\rm d}E{\rm d}\Omega=5.8^{+10.1}_{-3.7}\times 10^{-8}~\si{GeV~cm^{-2}~s^{-1}~sr^{-1}}$. If such a neutrino originated from the evaporation of a PBH, the implied flux would be significantly larger than the diffuse flux inferred from \textsc{IceCube} observations at lower energies. As shown in Fig.~\ref{fig:fluxes}, the expected PBH-induced flux is energetically broad nearby the primary spectrum peak, making it difficult to reconcile with such a high-energy detection unless the event is an extremely rare statistical fluctuation~\cite{Boccia:2025hpm}. 

Ref.~\cite{Boccia:2025hpm} considered PBHs with a constant emission rate, effectively considering the \textit{fast} decay scenario. Given the energies involved, we expect such signal to be sourced by a PBH of mass around $10^{6}\,\rm g$ whose flux, as already discussed above, is primarily constrained by the \textit{background-inclusive} analysis of the diffuse \textsc{Lhaaso} dataset.
If instead the signal originates from memory-burdened PBHs transitioning today, or from the mergers of such PBHs, then the cosmological component of the neutrino flux could be significantly enhanced. Since the measured flux will not strengthen the constraint, we do not consider it in our analysis. A detailed investigation of these possibilities is beyond the scope of this work and is left for future studies.

\subsection{Cosmological probes}
\label{subsec:cosmoprobes}
In this section, we examine additional complementary constraints derived from cosmological probes. In general, evaporating PBHs can inject highly energetic particles into the photon-baryon fluid at various cosmological epochs. The timing of this injection is governed by the PBH mass, which determines the characteristic timescale of the evaporation process. These injected particles trigger EM cascades that can interfere with the formation of light nuclei during BBN and distort the energy spectrum of the CMB. Such processes yield independent and complementary observational constraints on the abundance of PBHs.

\paragraph{BBN:}
Evaporating PBHs with masses $M_{\rm PBH}$ in the range $ 10^{10}~\mathrm{g} $ to $10^{13}~\mathrm{g}$ inject energy during or shortly after the formation of light elements altering the neutron-to-proton ratio and triggering photo-dissociation and hadro-dissociation of nuclei. Consequently, any deviation from the standard BBN scenario is subject to stringent constraints~\cite{Carr:2009jm, Carr:2020gox, Haque:2024eyh, Keith:2020jww, Boccia:2024nly}. For the merger case, the energy contribution from PBHs that merge and resume their Hawking evaporation through BBN is negligible due to the constraining power of CMB. In fact, the merger case redshifts approximately in the same way as for slowly transitioning PBHs, for which Refs.~\cite{Montefalcone:2025akm,Dvali:2025ktz} showed that BBN constraints are always subdominant. For this reason, BBN bounds are not showed in the left panel of Fig.~\ref{fig:money_plot_decay_all}. However, for $q \lesssim 1$, the bounds derived in Ref.~\cite{Thoss:2024hsr} due to the semiclassical decay of these objects still apply and are therefore shown in the left panel of Fig.~\ref{fig:money_plot_merger_all} with the label $\rm BBN_{SC}$. In the right panel of the same Figure, 
these constraints, which we did not re-derive, could be leading in the mass window between $10^{10}-10^{13}\,\rm g$. 
Note that for $q \lesssim 10^{-12}$, PBHs with $M_{\rm PBH}\lesssim10^{13}\si{g}$ can enter the memory burden phase before $1\,\mathrm{s}$, evading BBN constraints.

\paragraph{CMB:} 
During the cosmic dark ages, PBHs can inject energetic electrons and photons into the interGalactic medium (IGM), resulting in distortions of the CMB energy spectrum with potentially observable consequences. Constraints on evaporating {memory-burdened} PBH, accounting for both their semiclassical evaporation phase and the residual {memory burden floor}, have been derived in~\cite{Thoss:2024hsr}. Furthermore, in scenarios featuring a slowly-developing memory burden, Refs.~\cite{Dvali:2025ktz,Montefalcone:2025akm} have recently obtained bounds by mapping the PBH population onto a decaying DM framework, following the approach introduced in~\cite{Keith:2020jww}, originally developed for semiclassical PBHs. Given the relative strength of these CMB-based constraints in comparison to those from indirect detection searches, we compute them explicitly below. As a novel contribution, we also extend the formalism to derive CMB limits in the case of PBH mergers.

\medskip
The redshift evolution of the energy deposition history from evaporating memory-burdened PBHs depends on the specific scenario under consideration,  $\mathcal{S} = \{\textit{decay}, \textit{merger}\}$. In a compact form, the energy density deposited in the plasma is given by
\begin{equation}
    \frac{\dd E}{\dd V\,\dd t}\bigg|^{\mathcal S}_{\rm dep} = p_{\rm PBH}^{\mathcal{S}}(z)\, \rho_{{\rm DM},0}(1+z)^3 \, ,
    \label{eq:dedvdtDepPBH}
\end{equation}
where \( p^{\mathcal{S}}_{\rm PBH}(z) \) contains all the information about the source and the efficiency with which the injected energy ionizes the gas. More specifically, the deposited power takes the form
\begin{equation}
    p^{\mathcal{S}}_{\rm PBH}(z) = f_{\rm ion}^{\mathcal{S}}(z) \left[\frac{f_{\rm PBH}}{\tau_{\rm SC}} \, \xi^{\mathcal{S}}(z) \frac{\delta M_{\rm PBH}^{\rm EM}}{M_{\rm PBH}} \right] \,,
    \label{eq:pcases}
\end{equation}
where the term in square brackets represents the injected power. Here, $\delta M^{\rm EM}_{\rm PBH}$ is the EM energy fraction of the PBH mass\footnote{In the PBH mass range under consideration, the EM mass fraction $\delta M_{\rm PBH}^{\rm EM} = \tau_{\rm SC} \int E\, dE ( dN/(dE\,dt)|_\gamma + dN/(dE\,dt)|_{e^\pm} )$ outputted by {\tt BlackHawk}
accounts for approximately 40\% of the total PBH mass.}, and the redshift-dependent suppression factor $\xi^{\mathcal{S}}(z)$ is defined in~\eqref{eq:cases}. In the \textit{fast} scenario, $\xi^{\mathcal{S}}(z)$ is time-independent, making this case analogous to standard decaying DM, with an energy injection rate that scales as $(1+z)^3$. In the \textit{slow} and \textit{merger} scenarios, the suppression parameters acquire additional time dependence, resulting in energy injection rates that scale approximately as $(1+z)^{9/2}$ and $(1+z)^5$, respectively, during the matter-dominated era, characterized by $H(z) = \Omega_{\rm DM} H_0 (1+z)^{3/2}$. In~\eqref{eq:pcases}, $f_{\rm ion}^{\mathcal{S}}(z)$ denotes the ionization efficiency function. We note that, within the redshift range of interest, other energy deposition channels—such as excitation and heating—are subdominant and can be neglected.
Following~\cite{Slatyer:2015jla, Slatyer:2015kla}, we compute the efficiency functions by extracting the numerical results available online at \href{http://nebel.rc.fas.harvard.edu/epsilon}{http://nebel.rc.fas.harvard.edu/epsilon}. To fix ideas with concrete values, if we consider the redshift at which the CMB is most sensitive to standard decaying DM, namely $z^*_{\rm dec} = 300$, we find that the corresponding ionization efficiency factors $f_{\rm ion}^{\mathcal{S}}(z^*_{\rm dec})$ for $M_{\rm PBH}\lesssim 10^{10}\si{g}$ are approximately $f^{\rm fast}_{\rm ion}(300)\simeq 0.02$ for the fast case, $f^{\rm slow}_{\rm ion}(300)\simeq 0.04$ for the slow case, and $f^{\rm merger}_{\rm ion}(300)\simeq 0.06
$ for the merger case, respectively.

\smallskip
Having established the expression for the deposited energy, we now present our analytical framework to compute the CMB limits on the parameter space of evaporating, memory-burdened PBHs. Specifically, we outline a procedure to rescale the \textsc{Planck} constraints on the lifetime of standard decaying DM. Our method relies on the fact that the CMB constraints are primarily governed by the redshift dependence of the visibility function, which quantifies the sensitivity of CMB anisotropies to exotic energy injection occurring after recombination. The procedure relies on two key assumption. First, we assume that the dominant impact on the CMB arises from energy injections occurring near a characteristic redshift $z^\star$ corresponding to the peak of the visibility function $W(z)\propto d\tau/dz~e^{-\tau(z)}$. This peak typically occurs after recombination and reflects the redshift at which exotic energy injection has the most pronounced effect on the CMB power spectrum. Second, in order to reliably rescale constraints from standard decaying DM scenarios to alternative exotic energy injection from evaporating memory-burdened PBHs, we require that the visibility function of the new model, evaluated at its post-recombination maximum, matches in magnitude that of the benchmark case. This condition ensures that the perturbation to the ionization history—and hence the impact on CMB anisotropies—is comparable, thereby justifying the rescaling of the original limits.

From the first assumption, one can straightforwardly derive an analytical expression for the free electron fraction at redshift $z^\star$ for any given soft energy injection model. The key quantity entering the visibility function $W(z)$ is the photon optical depth, which is sensitive to variations in the free electron fraction $x_e(z)$, and scales with redshift as $\dd \tau(z)/\dd z \propto x_e(z) (1+z)^2 / H(z)$. Any additional energy injection modifies the evolution of $x_e(z)$, thereby altering the optical depth. Provided that the energy injection rate does not vary too rapidly with time, there exists a characteristic redshift $z^\star$ (after CMB decoupling) at which the visibility function is maximized. After decoupling, the free electron fraction rapidly decreases, and the optical depth becomes much smaller than unity. Therefore, the condition for maximizing the visibility function, $\dd W(z)/\dd z|_{z=z^\star} = 0$, reduces to the simpler requirement that $\dd x_e(z)/\dd z|_{z=z^\star} \approx 0$. From this condition, one can immediately read off from the system of coupled differential equations that describe the evolution of the free electron fraction $x_e(z)$ and the IGM temperature $T_{\rm IGM}(z)$ (see, for instance,~\cite{DAmico:2018sxd, Giesen:2012rp} and references therein), the free electron fraction at $z^\star$ that yields
\begin{equation}
     x_e(z^\star) \approx \left[\frac{1}{{\cal P}_2(z^\star)E_0 n_{\rm H,0}^2 (1+z^\star)^6}\frac{1}{\alpha_{\rm H}(T_{\rm IGM}(z^\star))}\frac{\dd E}{\dd V \dd t}\bigg|_{\rm dep}\hspace{-.2cm}(z^\star)\right]^{1/2} \ ,
     \label{eq:dxedz}    
\end{equation}
where $\alpha_{\rm H}(T_{\rm IGM})$ is the hydrogen recombination coefficient, $n_{\rm H,0}$ is the present-day hydrogen number density, $E_0$ is the ionization threshold energy, and ${\cal P}_2(z^\star)$ is the probability for an electron in the $n=2$ state to reach the ground state before being ionized~\cite{Giesen:2012rp}. This probability is close to unity after recombination. We stress again that~\eqref{eq:dxedz} is valid in the regime $x_e \ll 1$. The IGM temperature $T_{\rm IGM}$ is only mildly affected because, for relatively small energy injection, electrons remain in thermal equilibrium with CMB photons for $z \gtrsim 130$, implying $T_{\rm IGM}(z) \simeq T_{\rm CMB}^0 (1+z)$.

\smallskip
Having at our disposal an analytic expression for the free electron population at the peak redshift, we can now proceed with the rescaling of the bound as already discussed above. The \textsc{Planck} constraints on annihilating and decaying DM, derived from distortions in the CMB temperature anisotropies, have been computed for all possible SM primary channels in several studies. As a benchmark, we consider the decay process $\chi \to e^+e^-$, for which the bound on the decay rate is $\Gamma_\chi \lesssim \Gamma^{90\%}_\chi = 10^{-24}\,\si{s}^{-1}$, assuming a DM mass of $m_\chi = 1\,\si{TeV}$.

This bound arises primarily from redshift $z_{\rm dec}^\star \sim 300$. Using this limit the  deposited energy density by decaying DM at the peak redshift is $f^{\chi\to ee}_{\rm ion}(z^*_{\rm dec}) \Gamma_\chi^{90\%} \rho_{\rm DM,0} (1+z^*_{\rm dec})^3$ with $f^{\chi\to ee}_{\rm ion}(z^*_{\rm dec})\simeq 0.02$ (see for example Fig.~1 of~\cite{Slatyer:2016qyl}). 
The last ingredient we need to determine is the redshift at which the window function peaks in the context of evaporating memory-burdened PBHs. 
For the \textit{fast} decay scenario, the redshift dependence is exactly the one of standard decaying DM and therefore adopt $z^*_{\rm fast} = 300$ as a representative value. For the \textit{slow} and \textit{merger} scenarios, the energy injection is steeper in redshift, as already discussed above, but softer than the case of standard annihilating DM (where the bound arises primarily from $z_{\rm ann}^\star\sim500-600$~\cite{Finkbeiner_2012,Slatyer:2016qyl}). More specifically, we choose these typical redshifts to be \( z_{\rm slow}^\star \sim 400 \), and \( z_{\rm merger}^\star \sim 450 \) for the \textit{slow}, and \textit{merger} scenarios, respectively. For completeness, we also report the values of $f_{\rm ion}^{\cal S}(z_{\cal S}^\star)$ for $M_{\rm PBH}\lesssim 10^{10}\si{g}$ in both the slow and merger scenarios: $f^{\rm slow}_{\rm ion}(400)\simeq0.05$ and $f^{\rm merger}_{\rm ion}(450)\simeq0.08$.
By using~\eqref{eq:dedvdtDepPBH},~\eqref{eq:pcases},~\eqref{eq:dxedz} and in a matter-dominated era, we impose the scaling condition commented in above:
\begin{equation}
    \left.\frac{d\tau_{\rm dec}}{dz}\right|_{z^\star_{\rm dec}} \approx \left.\frac{d\tau_{\mathcal S}}{dz}\right|_{z^\star_{\cal S}} \ ,~ \mbox{i.e.~}~p_{\rm PBH}^{\cal S}(z^\star_{\cal S})\approx\frac{\alpha_{\rm H}(T_{\rm IGM}(z^\star_{\cal S}))}{\alpha_{\rm H}(T_{\rm IGM}(z^\star_{\rm dec}))}\frac{(1+z^\star_{\cal S})^2}{(1+z^\star_{\rm dec})^2}\Gamma_\chi^{90\%}f_{\rm ion}^{\chi\to ee}(z^\star_{\rm dec}) \ ,
\end{equation}
where $\alpha_{\rm H}(T_{\rm IGM}(z^\star_{\cal S}))/\alpha_{\rm H}(T_{\rm IGM}(z^\star_{\rm dec}))\approx\left[(1+z_{\rm dec})/(1+z^\star_{\cal S})\right]^{0.68}$, in the redshift range between 100 and 1000. As a validation of our procedure, we successfully reproduce the well-known results for the case of annihilating DM. In this case, the deposited power takes the form $f_{\rm ion}^{\rm ann}(z^\star_{\rm ann})\cdot\rho_{\rm DM,0}(1+z^\star _{\rm ann})^3\cdot \langle\sigma v\rangle/m_\chi$, where $f_{\rm ion}^{\rm ann}(z^\star_{\rm ann})\sim 0.15$. Furthermore, we recover the CMB bounds associated with the semiclassical evaporation phase, previously computed in~\cite{Thoss:2024hsr}. These bounds were obtained using dedicated numerical codes and a significantly more sophisticated analysis, which goes beyond the scope of the present work.

\medskip
We present the results for the \textit{decay} and \textit{merger} scenarios as red solid lines in Figs.~\ref{fig:money_plot_decay_all} and~\ref{fig:money_plot_merger_all}, respectively. In both cases, CMB constraints are competitive, and in certain regions of parameter space even comparable, to those from indirect detection probes. As shown in the left panel of Fig.~\ref{fig:money_plot_merger_all}, the CMB constraint probes down to $f_{\rm PBH} \simeq 10^{-3} - 10^{-2}$ for a mass fraction $q = 0.5$, while it excludes down to $q \simeq 10^{-3} - 10^{-2}$ assuming $f_{\rm PBH} = 1$. The results for the \textit{slow} decay scenario are displayed in the left panel of Fig.~\ref{fig:money_plot_decay_all}, where the CMB constraint reaches down to $f_{\rm PBH} \, \delta \simeq 10^{-10}$ for PBH masses below $10^{13}\,\mathrm{g}$. For larger masses, the CMB constraint becomes the leading bound and naturally aligns with the semiclassical limit derived in~\cite{Thoss:2024hsr}. In the case of the \textit{fast} decay scenario, the constraint extends over the entire PBH mass range considered in this work, i.e., $M_{\rm PBH} = (10$ – $10^{15})\,\mathrm{g}$, and reaches the benchmark value $k = 2$ at  $M_{\rm PBH} \simeq 2 \times 10^4\,\si{g}$.

\subsection{Summary of constraints and implications for memory burden}\label{subsec:summaryboundsandtheory}

For phenomenological purposes, we treated $q$ and $\delta$ as independent parameters and performed a scan over their respective ranges. However, as discussed in Sec.~\ref{sec:essence}, if one adopts the simplified prototype Hamiltonian~\eqref{eq:htot} as a fundamental description of PBHs approaching the memory burden regime, then both $q$ and $\delta$ are uniquely determined by the critical exponent $p$, as given in~\eqref{eq:deltamovm} and~\eqref{eq:deltapmapping}. Naturally, additional effects not captured in the prototype Hamiltonian~\eqref{eq:htot} may be present, so this identification should be taken with caution.
Below we summarize the main results of our parameter space analysis and clarify the mapping to the critical exponent $p$.

\paragraph{Semiclassical constraints.} Before the introduction of memory burden, pre-existing constraints due to PBHs lighter than the asteroid mass window - $M_{\rm PBH}\lesssim 10^{17}\,\rm g$ - follows from measurements of Galactic and extragalactic $\gamma$-rays, CMB and BBN. This is relevant for PBHs as light as $10^{10}\,\rm g$. In the case of memory burdened BHs, evading these constraints requires $q\ll 1$ (for $q\lesssim\mathcal{O}(1)$ similar constraints emerge almost unaltered~\cite{Thoss:2024hsr}, and are denoted with the pedix SC in the left panel of Fig.~\ref{fig:money_plot_merger_all}). These constraints are fully lifted if PBHs enter the memory burden phase before BBN era, requiring $q\, \tau_{\rm SC}\lesssim \mathcal{O}({\rm s})$, implying $q\lesssim \mathcal{O}(10^{-23})$ for a PBH of mass $10^{17}\,\rm g$. Notice that this is of order $q\sim S^{-1/2}$. In fact, for such a functional scaling of $q$, none of the semiclassical constraints apply to memory burden BHs. Incidentally, this corresponds to the critical exponent characterizing the prototype Hamiltonian $p=2$ in \eqref{eq:htot}. For larger value of $p$, pushing $q$ towards one, these constraints are present and force memory burden PBHs to be lighter than $10^{10}\,\rm g$.

\paragraph{Merger.} The merger case relies on the validity of the  contribution to the merger rate due to binary formed in the early Universe c.f.,~\eqref{eq:mergerrate}. We showed that the constraints resulting from \textit{merger} rules out memory burdened PBHs lighter than $10^{10}\,\rm g$ unless $q\lesssim 10^{-2}-10^{-3}$ (see Fig.~\ref{fig:money_plot_merger_all}). The leading constraints are obtained from \textsc{Auger} in the mass range $10$--$10^{4}\,\si{g}$, \textsc{Kascade-Grande} for $10^{4}$--$10^{5}\,\si{g}$, \textsc{Lhaaso} for $10^{5}$--$10^{8}\,\si{g}$, and \textsc{Fermi} for $10^{8}$--$10^{10}\,\si{g}$. These conclusions follow from a \textit{background-inclusive} analysis of the \textsc{Lhaaso} and \textsc{Fermi} datasets, which enhances the sensitivity of $\gamma$-ray observations compared to neutrino constraints.
As discussed above, for $M_{\rm PBH} \gtrsim 10^{10}\,\si{g}$, semiclassical bounds provide the most stringent constraints.
 
In terms of critical exponent $p$ our results suggests $p\lesssim 8$ and $p\lesssim 4$ for PBHs of mass $10^{10}\rm g$ and $10^{3}\rm g$ respectively. In our analysis we conservatively included suppression factors in the merger rate \eqref{eq:mergerrate} - contrary e.g., to~\cite{Kohri:2024qpd} where the gravitational counterpart of the merger of memory burden PBHs had been analyzed. To present day, these merger rates are estimates extrapolated from studies applied to heavier PBHs of solar mass order~\cite{Raidal:2024bmm} and around the asteroid mass window~\cite{Franciolini:2022htd}. 
Therefore, in the future we plan to explore the merger rate of ultralight PBHs in detail, bearing in mind that any increase in this quantity can strengthen these constraints.

\begin{figure*}[t!]
    \centering
        \begin{subfigure}[b]{0.49\textwidth}
        \centering
        \includegraphics[width=\linewidth]{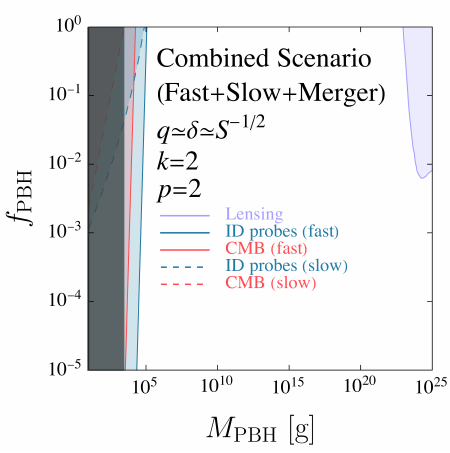}
    \end{subfigure}
    \,
    \begin{subfigure}[b]{0.485\textwidth}
        \centering
        \includegraphics[width=\linewidth]{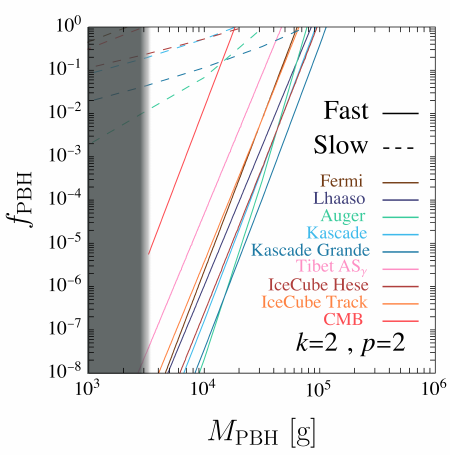}
    \end{subfigure}
    \caption{Constraints for the combined scenario (\textit{fast}+\textit{slow}+\textit{merger}) in the $(M_{\rm PBH},f_{\rm PBH})$ plane in the $k=2$, $p=2$ benchmark, i.e., $q \simeq \delta \simeq S^{-1/2}$ c.f., \eqref{eq:deltamovm} and \eqref{eq:deltapmapping}. Solid and dashed lines correspond to the \textit{fast} and \textit{decay} scenarios, respectively, shown for both indirect detection experiments (color scheme is chosen as in Figs.~\ref{fig:money_plot_merger_all} and \ref{fig:money_plot_decay_all}) and CMB limits (red lines). In the left panel, lensing constraints (see e.g.,~\cite{Carr:2020gox}) at the high-mass end of the window are shown as a light purple solid line. The right panel shows the same as the left one, but zoomed into the PBH mass window from $10^3\si{g}$ to $10^6\si{g}$. In both panels, the shaded region corresponds to the part of the parameter space for which $\tau \lesssim t_0$ c.f., \eqref{tMB}. Notice that the constraints from the width of the transition (slow case), are unaffected by different choices of $k$.}
    \label{fig:fastslow}
\end{figure*}

\paragraph{Fast decay.} 
We revisited existing bounds on evaporating PBHs in the memory burden case. Due to the exponential dependence of the suppression power, given by $S^{-k}$, the constraints are quite steep in mass, and very sensible to the parameter $k$ as well. For the case $k=2$, motivated by numerical and analytical studies of memory burden~\cite{Dvali:2020wft,Dvali:2024hsb}, the bounds are shown in the right panel of Fig.~\ref{fig:money_plot_decay_all} and in Fig.~\ref{fig:fastslow}, implying $M_{\rm PBH}\gtrsim 10^5\,\rm g$ for these objects to constitute the entirety of DM. Larger values of $k$ lower this mass threshold and, in general, for $f_{\rm PBH}=1$, a scan of the parameter $k$ is shown in the right panel of Fig.~\ref{fig:money_plot_decay_all}. 

\paragraph{Slow decay. } The bounds from the width of the transition are summarized in the left panel of Fig.~\ref{fig:money_plot_decay_all}. Notice that due to the approximate analytic scaling \eqref{kappafullmem} the combination $f_{\rm PBH}\,\delta$ is the one constrained. For $f_{\rm PBH}=1$, $\delta\lesssim 10^{-10}-10^{-13}$ in the mass window of interest. Our findings indicate that the leading experiments are \textsc{Auger} in the mass range $10$--$10^{4}\,\si{g}$, \textsc{Kascade-Grande} for $10^{4}$--$10^{6}\,\si{g}$, \textsc{Lhaaso} for $10^{6}$--$10^{9}\,\si{g}$, and \textsc{Fermi} for $10^{9}$--$10^{11}\,\si{g}$ (both including background). For higher masses, CMB bounds become relevant. We remark that, in the mass window relevant for \textsc{Lhaaso} and partially for \textsc{Fermi}, the \textit{background-inclusive} analysis leads to much more stringent $\gamma$-ray bounds compared to those derived from neutrino data and CMB anisotropies.

In the same figure, dashed lines show the scaling of $\delta$ for $p=2$ and $p=3$ from which we infer that for the former (latter) case, PBHs of mass $M_{\rm PBH}\gtrsim 10^{5}\,g$ ($M_{\rm PBH}\gtrsim 10^{11}\,\rm g$) can be the DM. Notice that $M_{\rm PBH}\gtrsim 10^{5}\,\rm g$ is also attained for $k=2$.

\paragraph{Combined scenario. }
The bounds on $\delta$ and $k$ lead to different steepness of the constraints as a function of mass, as shown in Fig.~\ref{fig:fastslow} implying that for $k\gtrsim 2$, the finite width of the transition always offers the leading constraint around $f_{\rm PBH}$. This can be inferred from the right panel of Fig.~\ref{fig:money_plot_decay_all} where for $k>2$ the bound shifts to lower masses. Moreover, independently of the value of $k$, all PBHs in the relevant mass window are still transitioning today as long as $p>2$, implying that the leading constraints stems from this, rather than from the memory burden ``floor''. In fact, the position of the $p=3$ line in the left panel of Fig.~\ref{fig:money_plot_decay_all} shows that PBHs of mass smaller than $10^{11}\,\rm g$ are still transitioning today. Differently stated, a PBH in the memory burden ``floor'' today is realized only in a corner of parameter space, suggesting that this case is more of an exception rather than the norm thereby implying that any probe of memory burden BH would likely follow from their transition phase.

For $p=3$, PBHs with mass $\gtrsim 10^{12}\,\rm g$ would end their semiclassical evaporation around $\mathcal{O}({\rm s})$ and are potentially subjected to constraints from BBN. This, when combined with the constraints on the width discussed in the previous paragraph, leaves but a small window parameter space, $10^{11}\,{\rm g}\lesssim M_{\rm PBH}\lesssim 10^{12}\,\rm g$, where PBHs can compose the entirety of DM. Finally, if we insist on characterizing $\delta$ and $q$ uniquely in terms of the critical exponent $p$ (i.e., $q\simeq \delta$ up to logarithmic factors), considerations from the transition width are more constraining than those from mergers - although the latter has the advantage of relying solely on the semiclassical evaporating phase. For $p=2$, all semiclassical constraints are lifted as well\footnote{Notice that there may be mild constraints around $ 10^{17}\,\rm{g} $ arising from BBN. In fact, the estimates required to evade them, as discussed in the context of semiclassical constraints above, appear to be borderline}, effectively leading to an extension of the asteroid mass window $10^{5}\,\si{g}\lesssim M_{\rm PBH}\lesssim 10^{23}\,\si{g}$ in which PBHs can constitute the entirety of the DM.

\section{Conclusion}\label{sec:conclusion}

The memory burden effect can be summarized in a single statement~\cite{Dvali:2018xpy,Dvali:2018ytn,Dvali:2020wft,Alexandre:2024nuo,Dvali:2024hsb}: 
\vspace{0.1em}

\begin{center}
\textit{The memory stored in a configuration resists its decay}.
\end{center}

\vspace{0.2em}

\noindent This suggests the possibility that evaporating PBHs might be stabilized due to quantum effects backreacting on the semiclassical dynamics. As a consequence, PBHs lighter than $10^{15}\rm g$, traditionally assumed to be too short-lived according to a naive extrapolation of Hawking rate through the entirety of the PBH lifetime, become viable DM candidates.

In this work, we constrained the viable parameter space for memory-burdened PBHs. The phenomenon is characterized by three parameters, $k,q$ and $\delta$ as nicely depicted in Fig.~\ref{fig:qualitative}. The first one has been subjected to several phenomenological studies in the literature~\cite{Thoss:2024hsr,Chianese:2024rsn,Liu:2025vpz,Chianese:2025wrk,Tan:2025vxp} and describes the memory burden ``floor'' i.e., the rate of emission in the full memory burden phase - characterized by a suppression of powers of $S^{-k}$. We recapped on these existing constraints clarifying some misunderstandings regarding the theoretical mapping between the fundamental parameters of the PBHs and the resulting flux. In particular, we assumed no mass tracking by the Hawking emission throughout the semiclassical phase, as pointed out in~\cite{Dvali:2025ktz}. Unless the BH enters the memory burden phase after shredding an $\mathcal{O}(1)$ fraction of its initial mass such an effect is non-negligible.

A novel aspect of this analysis is the full characterization of the parameter space according to two necessary features of memory-burdened PBHs. The first is that these objects undergo mergers in the present Universe, producing semiclassical “young” BHs that resume Hawking evaporation at an unsuppressed rate, as pointed out in~\cite{Zantedeschi:2024ram}. This constrains the duration of the semiclassical phase, parametrized by the mass fraction emitted in the semiclassical phase given by $q$. The second point, recently emphasized in~\cite{Dvali:2025ktz}, is that the transition into the memory burden phase cannot be instantaneous and may instead occur over cosmological timescales. This transition is characterized by the width $\delta$. 

Considering each single scenario separately, the viable parameter spaces for the three model-independent quantities $k,q$ and $\delta$ are nicely summarized in Figs.~\ref{fig:money_plot_merger_all} and~\ref{fig:money_plot_decay_all}.
We have performed, for the first time, a \textit{background-inclusive} analysis for both the \textsc{Fermi} and \textsc{Lhaaso} experiments. As a result, we find that the constraints from $\gamma$-rays are more stringent than those derived from neutrino observations, where the inclusion of a background component has a negligible impact due to the limited number of data points and their large statistical uncertainties. The only exception is for PBHs of mass $\sim 10^{5}- 10^{7}\,$g where neutrinos offer competitive limits. This work thus provides the first comprehensive and comparative study across different indirect detection experiments in all memory burden scenarios. In particular, we present the first dedicated analysis for both the \textit{slow} and \textit{merger} scenarios, while also improving the existing analysis for the \textit{fast} decay case by avoiding semiclassical mass tracking.
Furthermore, for the first time, we derive the CMB constraints semi-analytically in the memory burden phase for the \textit{merger} and \textit{slow} scenarios (some estimates for this case have been provided previously in~\cite{Dvali:2025ktz} and in the more technical analysis of \cite{Montefalcone:2025akm}), recasting these two memory burden PBH scenarios into the standard decaying DM framework, and successfully matching existing semiclassical results obtained via numerical codes.

From the theoretical point of view, the parameter $k$ is expected to be an integer of $\mathcal{O}(1)$, with different studies favoring the specific value $k=2$~\cite{Dvali:2020wft,Dvali:2024hsb}. Notice that, however, only estimates of this quantity are provided due to the absence of a concrete model describing the fully quantum phase. 
On the contrary, $q$ and $\delta$ can be connected by the critical exponent $p$ entering in the prototype Hamiltonian \eqref{eq:htot}, via \eqref{eq:deltamovm} and \eqref{eq:deltapmapping}. As widely discussed in Sec.~\ref{sec:essence}, this is possible since $p$ effectively characterizes the potential around the semiclassical region, thereby describing how the backreaction energetically slows down the evaporating process. 
 Unfortunately, the actual value of the critical exponent $p$ is not known for the case of BH. However, if we insist on the parametrization in terms of the critical exponent which relates the parameters $q$ and $\delta$, the obtained constraints imply that memory burdened PBHs become viable DM as long as $p\lesssim 4$. For example, if $p=3$, $M_{\rm PBH}\simeq \mathcal{O}(10^{12}\,\rm g)$ become viable. As shown in Fig.~\ref{fig:fastslow}, a larger window is obtained if $p=2$ ($q\simeq \delta \simeq S^{-1/2}$ up to logarithmic corrections), for which PBHs can compose the entirety of DM in the mass range $10^5\,\si{g}\lesssim M_{\rm PBH}\lesssim 10^{23}\,\si{g}$ resisting their decay.

\section{Acknowledgments}
We thank Pasquale Serpico and Antonio Marinelli for useful discussion.
MZ acknowledges  the Laboratoire de Physique Théorique et Hautes Energies (LPTHE) for warm hospitality and useful discussion during the completion of this work. MZ is also thankful to Marco Calzà for useful discussion regarding {\tt BlackHawk} and to Gia Dvali and Sebastian Zell for useful ongoing discussion on memory burden phenomenon. The research conducted by AD, GM and PP receives partial funding from the European Union–Next generation EU (through Progetti di Ricerca di Interesse Nazionale (PRIN) Grant No. 202289JEW4).

\appendix
\section{{\tt BlackHawk} Clarifications}
\label{app:AppBlackHawkclar}

\begin{figure*}[t!]
    \centering
        \begin{subfigure}[b]{0.48\textwidth}
        \centering
        \includegraphics[width=\linewidth]{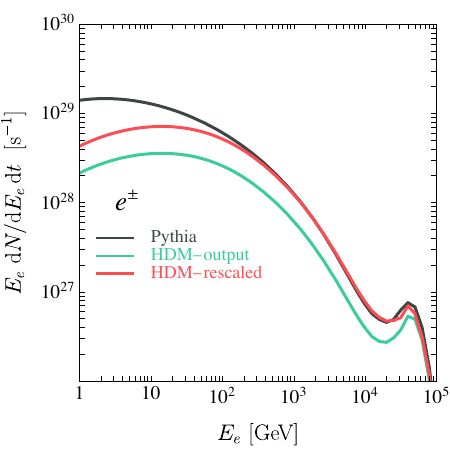}
    \end{subfigure}
   \,\,
    \begin{subfigure}[b]{0.48\textwidth}
        \centering
        \includegraphics[width=\linewidth]{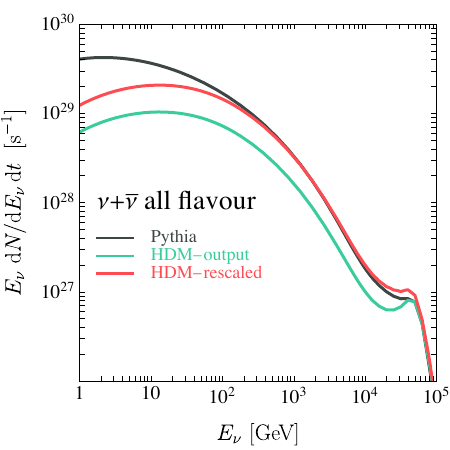}
    \end{subfigure}

    \vspace{0.2cm}
    \begin{subfigure}[b]{0.48\textwidth}
        \centering
        \includegraphics[width=\linewidth]{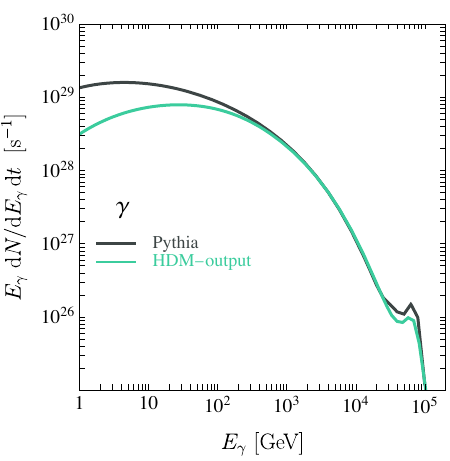}
    \end{subfigure}
    \,\,
    \begin{subfigure}[b]{0.48\textwidth}
        \centering
        \includegraphics[width=\linewidth]{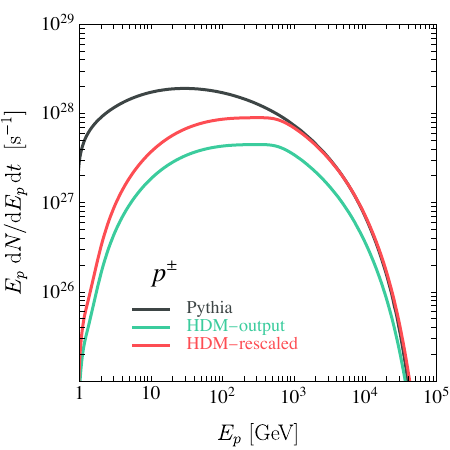}
    \end{subfigure}

    \caption{{\tt BlackHawk} clarifications. The spectra are computed for a representative example with $M_{\rm PBH}=10^9$ grams. Here we show that the {\tt BlackHawk} output for HDM differs by the one given by {\tt Pythia} by the relation in~\eqref{eq:BlackHawRescaling} for all the species that has an antiparticle.}
    \label{fig:BlackHawkClarification}
\end{figure*}

In this appendix, we clarify some details regarding the use of {\tt BlackHawk 2.3}, which is commonly employed in the literature to compute the initial particle emission rate. 
There are several options available for computing the secondary emission. In particular, we point out a potentially misleading interpretation of the final particle spectra when using the {\tt Pythia} and {\tt HDM} options.
When the output is set to {\tt Pythia}, the code returns a spectrum $\dd N/\dd E\,\dd t$ that refers to particle pairs. Conversely, when the {\tt HDM} option is selected, the output corresponds to the single-particle spectrum.

To make the {\tt HDM} output consistent with the {\tt Pythia} case, we apply the following rescaling:
\begin{equation}
    \left(\frac{\dd^2 N}{\dd E\,\dd t}\right)_{\rm sec}^{\rm HDM} = 
    2\left(\frac{\dd^2 N}{\dd E\,\dd t}\right)_{\rm sec}^{\rm HDM,\,output} - 
    \left(\frac{\dd^2 N}{\dd E\,\dd t}\right)_{\rm prim}^{\rm HDM,\,output},
    \label{eq:BlackHawRescaling}
\end{equation}
where we multiply the secondary emission by a factor of 2 and subtract the primary emission in order to avoid double counting.
This comparison is shown in Fig.~\ref{fig:BlackHawkClarification}, where the black, red, and green solid lines correspond to the {\tt Pythia} output, the unrescaled {\tt HDM} output, and the rescaled {\tt HDM} spectrum, respectively. 
With this cross-check, we confirm that the spectra computed with {\tt HDM} and {\tt Pythia} agree for $E/\Lambda(M_{\rm PBH}) \gtrsim 10^{-3}$. At lower values of this ratio, the two methods begin to diverge, as the hadronization processes are handled differently in each approach. 



\bibliographystyle{JHEP}
{\footnotesize
\bibliography{main}}

\providecommand{\href}[2]{#2}\begingroup\raggedright\begin{thebibliography}{100}

\bibitem{Zeldovich:1967lct}
Y.~B. Zel'dovich and I.~D. Novikov {\em Sov. Astron.} {\bf 10} (1967) 602.

\bibitem{Hawking:1971ei}
S.~Hawking {\em Mon. Not. Roy. Astron. Soc.} {\bf 152} (1971) 75.

\bibitem{Carr:1974nx}
B.~J. Carr and S.~W. Hawking {\em Mon. Not. Roy. Astron. Soc.} {\bf 168} (1974) 399--415.

\bibitem{Chapline:1975ojl}
G.~F. Chapline {\em Nature} {\bf 253} (1975), no.~5489 251--252.

\bibitem{Carr:1975qj}
B.~J. Carr {\em Astrophys. J.} {\bf 201} (1975) 1--19.

\bibitem{Carr:2020xqk}
B.~Carr and F.~Kuhnel {\em Ann. Rev. Nucl. Part. Sci.} {\bf 70} (2020) 355--394, [\href{http://arxiv.org/abs/2006.02838}{{\tt arXiv:2006.02838}}].

\bibitem{Green:2020jor}
A.~M. Green and B.~J. Kavanagh {\em J. Phys. G} {\bf 48} (2021), no.~4 043001, [\href{http://arxiv.org/abs/2007.10722}{{\tt arXiv:2007.10722}}].

\bibitem{Escriva:2022duf}
A.~Escriv\`a, F.~Kuhnel, and Y.~Tada \href{http://arxiv.org/abs/2211.05767}{{\tt arXiv:2211.05767}}.

\bibitem{Hawking:1975vcx}
S.~W. Hawking {\em Commun. Math. Phys.} {\bf 43} (1975) 199--220. [Erratum: Commun.Math.Phys. 46, 206 (1976)].

\bibitem{Bekenstein:1973ur}
J.~D. Bekenstein {\em Phys. Rev. D} {\bf 7} (1973) 2333--2346.

\bibitem{Carr:2009jm}
B.~J. Carr, K.~Kohri, Y.~Sendouda, and J.~Yokoyama {\em Phys. Rev. D} {\bf 81} (2010) 104019, [\href{http://arxiv.org/abs/0912.5297}{{\tt arXiv:0912.5297}}].

\bibitem{Carr:2016hva}
B.~J. Carr, K.~Kohri, Y.~Sendouda, and J.~Yokoyama {\em Phys. Rev. D} {\bf 94} (2016), no.~4 044029, [\href{http://arxiv.org/abs/1604.05349}{{\tt arXiv:1604.05349}}].

\bibitem{Carr:2020gox}
B.~Carr, K.~Kohri, Y.~Sendouda, and J.~Yokoyama {\em Rept. Prog. Phys.} {\bf 84} (2021), no.~11 116902, [\href{http://arxiv.org/abs/2002.12778}{{\tt arXiv:2002.12778}}].

\bibitem{Arbey:2019vqx}
A.~Arbey, J.~Auffinger, and J.~Silk {\em Phys. Rev. D} {\bf 101} (2020), no.~2 023010, [\href{http://arxiv.org/abs/1906.04750}{{\tt arXiv:1906.04750}}].

\bibitem{Boccia:2024nly}
A.~Boccia, F.~Iocco, and L.~Visinelli \href{http://arxiv.org/abs/2405.18493}{{\tt arXiv:2405.18493}}.

\bibitem{Dvali:2018xpy}
G.~Dvali \href{http://arxiv.org/abs/1810.02336}{{\tt arXiv:1810.02336}}.

\bibitem{Dvali:2018ytn}
G.~Dvali, L.~Eisemann, M.~Michel, and S.~Zell {\em JCAP} {\bf 03} (2019) 010, [\href{http://arxiv.org/abs/1812.08749}{{\tt arXiv:1812.08749}}].

\bibitem{Dvali:2020wft}
G.~Dvali, L.~Eisemann, M.~Michel, and S.~Zell {\em Phys. Rev. D} {\bf 102} (2020), no.~10 103523, [\href{http://arxiv.org/abs/2006.00011}{{\tt arXiv:2006.00011}}].

\bibitem{Alexandre:2024nuo}
A.~Alexandre, G.~Dvali, and E.~Koutsangelas {\em Phys. Rev. D} {\bf 110} (2024), no.~3 036004, [\href{http://arxiv.org/abs/2402.14069}{{\tt arXiv:2402.14069}}].

\bibitem{Dvali:2024hsb}
G.~Dvali, J.~S. Valbuena-Berm\'udez, and M.~Zantedeschi {\em Phys. Rev. D} {\bf 110} (2024), no.~5 056029, [\href{http://arxiv.org/abs/2405.13117}{{\tt arXiv:2405.13117}}].

\bibitem{Dvali:2021tez}
G.~Dvali, O.~Kaikov, and J.~S.~V. Berm\'udez {\em Phys. Rev. D} {\bf 105} (2022), no.~5 056013, [\href{http://arxiv.org/abs/2112.00551}{{\tt arXiv:2112.00551}}].

\bibitem{Dvali:2021bsy}
G.~Dvali \href{http://arxiv.org/abs/2103.15668}{{\tt arXiv:2103.15668}}.

\bibitem{Dvali:2023qlk}
G.~Dvali, O.~Kaikov, F.~K\"uhnel, J.~S. Valbuena-Bermudez, and M.~Zantedeschi {\em Phys. Rev. Lett.} {\bf 132} (2024), no.~15 151402, [\href{http://arxiv.org/abs/2310.02288}{{\tt arXiv:2310.02288}}].

\bibitem{Dvali:2020wqi}
G.~Dvali {\em JHEP} {\bf 03} (2021) 126, [\href{http://arxiv.org/abs/2003.05546}{{\tt arXiv:2003.05546}}].

\bibitem{Dvali:2019jjw}
G.~Dvali {\em Fortsch. Phys.} {\bf 69} (2021), no.~1 2000090, [\href{http://arxiv.org/abs/1906.03530}{{\tt arXiv:1906.03530}}].

\bibitem{Dvali:2019ulr}
G.~Dvali {\em Fortsch. Phys.} {\bf 69} (2021), no.~1 2000091, [\href{http://arxiv.org/abs/1907.07332}{{\tt arXiv:1907.07332}}].

\bibitem{Dvali:2021jto}
G.~Dvali {\em Phil. Trans. A. Math. Phys. Eng. Sci.} {\bf 380} (2021), no.~2216 20210071, [\href{http://arxiv.org/abs/2107.10616}{{\tt arXiv:2107.10616}}].

\bibitem{Dvali:2021ooc}
G.~Dvali and R.~Venugopalan {\em Phys. Rev. D} {\bf 105} (2022), no.~5 056026, [\href{http://arxiv.org/abs/2106.11989}{{\tt arXiv:2106.11989}}].

\bibitem{Dvali:2021rlf}
G.~Dvali and O.~Sakhelashvili {\em Phys. Rev. D} {\bf 105} (2022), no.~6 065014, [\href{http://arxiv.org/abs/2111.03620}{{\tt arXiv:2111.03620}}].

\bibitem{Dvali:2021ofp}
G.~Dvali, F.~K\"uhnel, and M.~Zantedeschi {\em Phys. Rev. Lett.} {\bf 129} (2022), no.~6 061302, [\href{http://arxiv.org/abs/2112.08354}{{\tt arXiv:2112.08354}}].

\bibitem{Zantedeschi:2022czs}
M.~Zantedeschi, {\em {On structure and primordial origin of black holes}}.
\newblock PhD thesis, Munich U., 2022.

\bibitem{Thoss:2024hsr}
V.~Thoss, A.~Burkert, and K.~Kohri {\em Mon. Not. Roy. Astron. Soc.} {\bf 532} (2024), no.~1 451--459, [\href{http://arxiv.org/abs/2402.17823}{{\tt arXiv:2402.17823}}].

\bibitem{Zantedeschi:2024ram}
M.~Zantedeschi and L.~Visinelli \href{http://arxiv.org/abs/2410.07037}{{\tt arXiv:2410.07037}}.

\bibitem{Dvali:2011aa}
G.~Dvali and C.~Gomez {\em Fortsch. Phys.} {\bf 61} (2013) 742--767, [\href{http://arxiv.org/abs/1112.3359}{{\tt arXiv:1112.3359}}].

\bibitem{Dvali:2012rt}
G.~Dvali and C.~Gomez {\em Phys. Lett. B} {\bf 719} (2013) 419--423, [\href{http://arxiv.org/abs/1203.6575}{{\tt arXiv:1203.6575}}].

\bibitem{Dvali:2012en}
G.~Dvali and C.~Gomez {\em Eur. Phys. J. C} {\bf 74} (2014) 2752, [\href{http://arxiv.org/abs/1207.4059}{{\tt arXiv:1207.4059}}].

\bibitem{Dvali:2012wq}
G.~Dvali and C.~Gomez \href{http://arxiv.org/abs/1212.0765}{{\tt arXiv:1212.0765}}.

\bibitem{Dvali:2013vxa}
G.~Dvali, D.~Flassig, C.~Gomez, A.~Pritzel, and N.~Wintergerst {\em Phys. Rev. D} {\bf 88} (2013), no.~12 124041, [\href{http://arxiv.org/abs/1307.3458}{{\tt arXiv:1307.3458}}].

\bibitem{Dvali:2015wca}
G.~Dvali and M.~Panchenko \href{http://arxiv.org/abs/1507.08952}{{\tt arXiv:1507.08952}}.

\bibitem{Michel:2023ydf}
M.~Michel and S.~Zell {\em Fortsch. Phys.} {\bf 71} (2023) 2300163, [\href{http://arxiv.org/abs/2306.09410}{{\tt arXiv:2306.09410}}].

\bibitem{Dvali:2025ktz}
G.~Dvali, M.~Zantedeschi, and S.~Zell \href{http://arxiv.org/abs/2503.21740}{{\tt arXiv:2503.21740}}.

\bibitem{Montefalcone:2025akm}
G.~Montefalcone, D.~Hooper, K.~Freese, C.~Kelso, F.~Kuhnel, and P.~Sandick \href{http://arxiv.org/abs/2503.21005}{{\tt arXiv:2503.21005}}.

\bibitem{IceCube:2018fhm}
{\bf IceCube} Collaboration, M.~G. Aartsen et~al. {\em Phys. Rev. D} {\bf 98} (2018), no.~6 062003, [\href{http://arxiv.org/abs/1807.01820}{{\tt arXiv:1807.01820}}].

\bibitem{IceCube:2020wum}
{\bf IceCube} Collaboration, R.~Abbasi et~al. {\em Phys. Rev. D} {\bf 104} (2021) 022002, [\href{http://arxiv.org/abs/2011.03545}{{\tt arXiv:2011.03545}}].

\bibitem{KM3NeT:2025npi}
{\bf KM3NeT} Collaboration, S.~Aiello et~al. {\em Nature} {\bf 638} (2025), no.~8050 376--382.

\bibitem{Airoldi:2025bgr}
L.~F.~T. Airoldi, G.~F.~S. Alves, Y.~F. Perez-Gonzalez, G.~M. Salla, and R.~Z. Funchal \href{http://arxiv.org/abs/2505.24652}{{\tt arXiv:2505.24652}}.

\bibitem{Airoldi:2025opo}
L.~F.~T. Airoldi, G.~F.~S. Alves, Y.~F. Perez-Gonzalez, G.~M. Salla, and R.~Z. Funchal \href{http://arxiv.org/abs/2505.24666}{{\tt arXiv:2505.24666}}.

\bibitem{Anchordoqui:2025xug}
L.~A. Anchordoqui, F.~Halzen, and D.~Lust \href{http://arxiv.org/abs/2505.23414}{{\tt arXiv:2505.23414}}.

\bibitem{Baker:2025cff}
M.~J. Baker, J.~Iguaz~Juan, A.~Symons, and A.~Thamm \href{http://arxiv.org/abs/2505.22722}{{\tt arXiv:2505.22722}}.

\bibitem{Klipfel:2025jql}
A.~P. Klipfel and D.~I. Kaiser \href{http://arxiv.org/abs/2503.19227}{{\tt arXiv:2503.19227}}.

\bibitem{Jiang:2025blz}
S.~Jiang and F.~P. Huang \href{http://arxiv.org/abs/2503.14332}{{\tt arXiv:2503.14332}}.

\bibitem{Boccia:2025hpm}
A.~Boccia and F.~Iocco \href{http://arxiv.org/abs/2502.19245}{{\tt arXiv:2502.19245}}.

\bibitem{Chianese:2024rsn}
M.~Chianese, A.~Boccia, F.~Iocco, G.~Miele, and N.~Saviano \href{http://arxiv.org/abs/2410.07604}{{\tt arXiv:2410.07604}}.

\bibitem{Arbey:2019mbc}
A.~Arbey and J.~Auffinger {\em Eur. Phys. J. C} {\bf 79} (2019), no.~8 693, [\href{http://arxiv.org/abs/1905.04268}{{\tt arXiv:1905.04268}}].

\bibitem{Arbey:2021mbl}
A.~Arbey and J.~Auffinger {\em Eur. Phys. J. C} {\bf 81} (2021) 910, [\href{http://arxiv.org/abs/2108.02737}{{\tt arXiv:2108.02737}}].

\bibitem{Bernal:2022swt}
N.~Bernal, V.~Mu\~noz Albornoz, S.~Palomares-Ruiz, and P.~Villanueva-Domingo {\em JCAP} {\bf 10} (2022) 068, [\href{http://arxiv.org/abs/2203.14979}{{\tt arXiv:2203.14979}}].

\bibitem{Bauer:2020jay}
C.~W. Bauer, N.~L. Rodd, and B.~R. Webber {\em JHEP} {\bf 06} (2021) 121, [\href{http://arxiv.org/abs/2007.15001}{{\tt arXiv:2007.15001}}].

\bibitem{DeRomeri:2024zqs}
V.~De~Romeri, Y.~F. Perez-Gonzalez, and A.~Tolino {\em JCAP} {\bf 04} (2025) 018, [\href{http://arxiv.org/abs/2405.00124}{{\tt arXiv:2405.00124}}].

\bibitem{Liu:2025vpz}
T.-C. Liu, B.-Y. Zhu, Y.-F. Liang, X.-S. Hu, and E.-W. Liang \href{http://arxiv.org/abs/2503.13192}{{\tt arXiv:2503.13192}}.

\bibitem{Chianese:2025wrk}
M.~Chianese \href{http://arxiv.org/abs/2504.03838}{{\tt arXiv:2504.03838}}.

\bibitem{Chaudhuri:2025rcs}
A.~Chaudhuri, K.~Pal, and R.~Mohanta \href{http://arxiv.org/abs/2505.09153}{{\tt arXiv:2505.09153}}.

\bibitem{Tan:2025vxp}
X.-h. Tan and Y.-f. Zhou \href{http://arxiv.org/abs/2505.19857}{{\tt arXiv:2505.19857}}.

\bibitem{Dvali:2021byy}
G.~Dvali, F.~K\"uhnel, and M.~Zantedeschi {\em Phys. Rev. D} {\bf 104} (2021), no.~12 123507, [\href{http://arxiv.org/abs/2108.09471}{{\tt arXiv:2108.09471}}].

\bibitem{Franciolini:2023osw}
G.~Franciolini and P.~Pani {\em Phys. Rev. D} {\bf 108} (2023), no.~8 083527, [\href{http://arxiv.org/abs/2304.13576}{{\tt arXiv:2304.13576}}].

\bibitem{Balaji:2024hpu}
S.~Balaji, G.~Dom\`enech, G.~Franciolini, A.~Ganz, and J.~Tr\"ankle {\em JCAP} {\bf 11} (2024) 026, [\href{http://arxiv.org/abs/2403.14309}{{\tt arXiv:2403.14309}}].

\bibitem{Haque:2024eyh}
M.~R. Haque, S.~Maity, D.~Maity, and Y.~Mambrini {\em JCAP} {\bf 07} (2024) 002, [\href{http://arxiv.org/abs/2404.16815}{{\tt arXiv:2404.16815}}].

\bibitem{Barman:2024iht}
B.~Barman, M.~R. Haque, and O.~Zapata {\em JCAP} {\bf 09} (2024) 020, [\href{http://arxiv.org/abs/2405.15858}{{\tt arXiv:2405.15858}}].

\bibitem{Bhaumik:2024qzd}
N.~Bhaumik, M.~R. Haque, R.~K. Jain, and M.~Lewicki {\em JHEP} {\bf 10} (2024) 142, [\href{http://arxiv.org/abs/2409.04436}{{\tt arXiv:2409.04436}}].

\bibitem{Barman:2024ufm}
B.~Barman, K.~Loho, and O.~Zapata {\em JCAP} {\bf 10} (2024) 065, [\href{http://arxiv.org/abs/2409.05953}{{\tt arXiv:2409.05953}}].

\bibitem{Kohri:2024qpd}
K.~Kohri, T.~Terada, and T.~T. Yanagida \href{http://arxiv.org/abs/2409.06365}{{\tt arXiv:2409.06365}}.

\bibitem{Borah:2024bcr}
D.~Borah and N.~Das {\em JCAP} {\bf 02} (2025) 031, [\href{http://arxiv.org/abs/2410.16403}{{\tt arXiv:2410.16403}}].

\bibitem{Barker:2024mpz}
W.~Barker, B.~Gladwyn, and S.~Zell \href{http://arxiv.org/abs/2410.11948}{{\tt arXiv:2410.11948}}.

\bibitem{Jiang:2024aju}
Y.~Jiang, C.~Yuan, C.-Z. Li, and Q.-G. Huang \href{http://arxiv.org/abs/2409.07976}{{\tt arXiv:2409.07976}}.

\bibitem{Loc:2024qbz}
N.~P.~D. Loc {\em Phys. Rev. D} {\bf 111} (2025), no.~2 023509, [\href{http://arxiv.org/abs/2410.17544}{{\tt arXiv:2410.17544}}].

\bibitem{Basumatary:2024uwo}
U.~Basumatary, N.~Raj, and A.~Ray {\em Phys. Rev. D} {\bf 111} (2025), no.~4 L041306, [\href{http://arxiv.org/abs/2410.22702}{{\tt arXiv:2410.22702}}].

\bibitem{Federico:2024fyt}
K.~Federico and S.~Profumo {\em Phys. Rev. D} {\bf 111} (2025), no.~6 063006, [\href{http://arxiv.org/abs/2411.17038}{{\tt arXiv:2411.17038}}].

\bibitem{Athron:2024fcj}
P.~Athron, M.~Chianese, S.~Datta, R.~Samanta, and N.~Saviano \href{http://arxiv.org/abs/2411.19286}{{\tt arXiv:2411.19286}}.

\bibitem{Barman:2024kfj}
B.~Barman, K.~Loho, and O.~Zapata {\em JCAP} {\bf 02} (2025) 052, [\href{http://arxiv.org/abs/2412.13254}{{\tt arXiv:2412.13254}}].

\bibitem{Bandyopadhyay:2025ast}
D.~Bandyopadhyay, D.~Borah, and N.~Das \href{http://arxiv.org/abs/2501.04076}{{\tt arXiv:2501.04076}}.

\bibitem{Calabrese:2025sfh}
R.~Calabrese, M.~Chianese, and N.~Saviano \href{http://arxiv.org/abs/2501.06298}{{\tt arXiv:2501.06298}}.

\bibitem{Dvali:2017nis}
G.~Dvali {\em Phys. Rev. D} {\bf 97} (2018), no.~10 105005, [\href{http://arxiv.org/abs/1712.02233}{{\tt arXiv:1712.02233}}].

\bibitem{Dvali:2018vvx}
G.~Dvali {\em Fortsch. Phys.} {\bf 66} (2018), no.~4 1800007, [\href{http://arxiv.org/abs/1801.03918}{{\tt arXiv:1801.03918}}].

\bibitem{Dvali:2018tqi}
G.~Dvali, M.~Michel, and S.~Zell {\em EPJ Quant. Technol.} {\bf 6} (2019) 1, [\href{http://arxiv.org/abs/1805.10292}{{\tt arXiv:1805.10292}}].

\bibitem{Averin:2016ybl}
A.~Averin, G.~Dvali, C.~Gomez, and D.~Lust {\em JHEP} {\bf 06} (2016) 088, [\href{http://arxiv.org/abs/1601.03725}{{\tt arXiv:1601.03725}}].

\bibitem{Averin:2016hhm}
A.~Averin, G.~Dvali, C.~Gomez, and D.~Lust {\em Mod. Phys. Lett. A} {\bf 31} (2016), no.~39 1630045, [\href{http://arxiv.org/abs/1606.06260}{{\tt arXiv:1606.06260}}].

\bibitem{Friedberg:1976me}
R.~Friedberg, T.~D. Lee, and A.~Sirlin {\em Phys. Rev. D} {\bf 13} (1976) 2739--2761.

\bibitem{Coleman:1985ki}
S.~R. Coleman {\em Nucl. Phys. B} {\bf 262} (1985), no.~2 263. [Addendum: Nucl.Phys.B 269, 744 (1986)].

\bibitem{Kusenko:1997vi}
A.~Kusenko, M.~E. Shaposhnikov, and P.~G. Tinyakov {\em Pisma Zh. Eksp. Teor. Fiz.} {\bf 67} (1998) 229, [\href{http://arxiv.org/abs/hep-th/9801041}{{\tt hep-th/9801041}}].

\bibitem{Ali-Haimoud:2017rtz}
Y.~Ali-Ha\"\i{}moud, E.~D. Kovetz, and M.~Kamionkowski {\em Phys. Rev. D} {\bf 96} (2017), no.~12 123523, [\href{http://arxiv.org/abs/1709.06576}{{\tt arXiv:1709.06576}}].

\bibitem{Raidal:2018bbj}
M.~Raidal, C.~Spethmann, V.~Vaskonen, and H.~Veerm\"ae {\em JCAP} {\bf 02} (2019) 018, [\href{http://arxiv.org/abs/1812.01930}{{\tt arXiv:1812.01930}}].

\bibitem{Liu:2018ess}
L.~Liu, Z.-K. Guo, and R.-G. Cai {\em Phys. Rev. D} {\bf 99} (2019), no.~6 063523, [\href{http://arxiv.org/abs/1812.05376}{{\tt arXiv:1812.05376}}].

\bibitem{Vaskonen:2019jpv}
V.~Vaskonen and H.~Veerm\"ae {\em Phys. Rev. D} {\bf 101} (2020), no.~4 043015, [\href{http://arxiv.org/abs/1908.09752}{{\tt arXiv:1908.09752}}].

\bibitem{Hutsi:2020sol}
G.~H\"utsi, M.~Raidal, V.~Vaskonen, and H.~Veerm\"ae {\em JCAP} {\bf 03} (2021) 068, [\href{http://arxiv.org/abs/2012.02786}{{\tt arXiv:2012.02786}}].

\bibitem{Jedamzik:2020ypm}
K.~Jedamzik {\em JCAP} {\bf 09} (2020) 022, [\href{http://arxiv.org/abs/2006.11172}{{\tt arXiv:2006.11172}}].

\bibitem{Young:2020scc}
S.~Young and A.~S. Hamers {\em JCAP} {\bf 10} (2020) 036, [\href{http://arxiv.org/abs/2006.15023}{{\tt arXiv:2006.15023}}].

\bibitem{Jedamzik:2020omx}
K.~Jedamzik {\em Phys. Rev. Lett.} {\bf 126} (2021), no.~5 051302, [\href{http://arxiv.org/abs/2007.03565}{{\tt arXiv:2007.03565}}].

\bibitem{Raidal:2024bmm}
M.~Raidal, V.~Vaskonen, and H.~Veerm\"ae, {\em {Formation of~Primordial Black Hole Binaries and~Their Merger Rates}}.
\newblock 2025.
\newblock \href{http://arxiv.org/abs/2404.08416}{{\tt arXiv:2404.08416}}.

\bibitem{Inman:2019wvr}
D.~Inman and Y.~Ali-Ha\"\i{}moud {\em Phys. Rev. D} {\bf 100} (2019), no.~8 083528, [\href{http://arxiv.org/abs/1907.08129}{{\tt arXiv:1907.08129}}].

\bibitem{Franciolini:2022htd}
G.~Franciolini, A.~Maharana, and F.~Muia {\em Phys. Rev. D} {\bf 106} (2022), no.~10 103520, [\href{http://arxiv.org/abs/2205.02153}{{\tt arXiv:2205.02153}}].

\bibitem{Sasaki:2018dmp}
M.~Sasaki, T.~Suyama, T.~Tanaka, and S.~Yokoyama {\em Class. Quant. Grav.} {\bf 35} (2018), no.~6 063001, [\href{http://arxiv.org/abs/1801.05235}{{\tt arXiv:1801.05235}}].

\bibitem{Raidal:2017mfl}
M.~Raidal, V.~Vaskonen, and H.~Veerm\"ae {\em JCAP} {\bf 09} (2017) 037, [\href{http://arxiv.org/abs/1707.01480}{{\tt arXiv:1707.01480}}].

\bibitem{DeLuca:2021hde}
V.~De~Luca, G.~Franciolini, P.~Pani, and A.~Riotto {\em JCAP} {\bf 11} (2021) 039, [\href{http://arxiv.org/abs/2106.13769}{{\tt arXiv:2106.13769}}].

\bibitem{DeLuca:2020bjf}
V.~De~Luca, G.~Franciolini, P.~Pani, and A.~Riotto {\em JCAP} {\bf 04} (2020) 052, [\href{http://arxiv.org/abs/2003.02778}{{\tt arXiv:2003.02778}}].

\bibitem{Page:1976ki}
D.~N. Page {\em Phys. Rev. D} {\bf 14} (1976) 3260--3273.

\bibitem{Page:1976df}
D.~N. Page {\em Phys. Rev. D} {\bf 13} (1976) 198--206.

\bibitem{Arbey:2025dnc}
A.~Arbey, M.~Calz\`a, and Y.~F. Perez-Gonzalez {\em Phys. Dark Univ.} {\bf 48} (2025) 101903, [\href{http://arxiv.org/abs/2502.17240}{{\tt arXiv:2502.17240}}].

\bibitem{Planck:2018vyg}
{\bf Planck} Collaboration, N.~Aghanim et~al. {\em Astron. Astrophys.} {\bf 641} (2020) A6, [\href{http://arxiv.org/abs/1807.06209}{{\tt arXiv:1807.06209}}]. [Erratum: Astron.Astrophys. 652, C4 (2021)].

\bibitem{Navarro:1996gj}
J.~F. Navarro, C.~S. Frenk, and S.~D.~M. White {\em Astrophys. J.} {\bf 490} (1997) 493--508, [\href{http://arxiv.org/abs/astro-ph/9611107}{{\tt astro-ph/9611107}}].

\bibitem{Riccardo_Catena_2010}
R.~Catena and P.~Ullio {\em Journal of Cosmology and Astroparticle Physics} {\bf 2010} (Aug., 2010) 004–004.

\bibitem{Salucci_2010}
P.~Salucci, F.~Nesti, G.~Gentile, and C.~Frigerio~Martins {\em Astronomy \&amp; Astrophysics} {\bf 523} (Nov., 2010) A83.

\bibitem{Iocco_2015}
F.~Iocco, M.~Pato, and G.~Bertone {\em Nature Physics} {\bf 11} (Feb., 2015) 245–248.

\bibitem{Esmaili_2015}
A.~Esmaili and P.~D. Serpico {\em Journal of Cosmology and Astroparticle Physics} {\bf 2015} (Oct., 2015) 014–014.

\bibitem{Blanco:2018bbf}
C.~Blanco {\em JCAP} {\bf 01} (2019) 013, [\href{http://arxiv.org/abs/1804.00005}{{\tt arXiv:1804.00005}}].

\bibitem{Capanema:2024nwe}
A.~Capanema and C.~Blanco {\em Comput. Phys. Commun.} {\bf 307} (2025) 109408, [\href{http://arxiv.org/abs/2408.03995}{{\tt arXiv:2408.03995}}].

\bibitem{Saldana-Lopez:2020qzx}
A.~Saldana-Lopez, A.~Dom\'\i{}nguez, P.~G. P\'erez-Gonz\'alez, J.~Finke, M.~Ajello, J.~R. Primack, V.~S. Paliya, and A.~Desai {\em Mon. Not. Roy. Astron. Soc.} {\bf 507} (2021), no.~4 5144--5160, [\href{http://arxiv.org/abs/2012.03035}{{\tt arXiv:2012.03035}}].

\bibitem{Cirelli:2010xx}
M.~Cirelli, G.~Corcella, A.~Hektor, G.~Hutsi, M.~Kadastik, P.~Panci, M.~Raidal, F.~Sala, and A.~Strumia {\em JCAP} {\bf 03} (2011) 051, [\href{http://arxiv.org/abs/1012.4515}{{\tt arXiv:1012.4515}}]. [Erratum: JCAP 10, E01 (2012)].

\bibitem{Slatyer:2009yq}
T.~R. Slatyer, N.~Padmanabhan, and D.~P. Finkbeiner {\em Phys. Rev. D} {\bf 80} (2009) 043526, [\href{http://arxiv.org/abs/0906.1197}{{\tt arXiv:0906.1197}}].

\bibitem{Fermi-LAT:2014ryh}
{\bf Fermi-LAT} Collaboration, M.~Ackermann et~al. {\em Astrophys. J.} {\bf 799} (2015) 86, [\href{http://arxiv.org/abs/1410.3696}{{\tt arXiv:1410.3696}}].

\bibitem{Abbasi:2021qfz}
R.~Abbasi et~al. {\em Astrophys. J.} {\bf 928} (2022), no.~1 50, [\href{http://arxiv.org/abs/2111.10299}{{\tt arXiv:2111.10299}}].

\bibitem{Arguelles:2019ouk}
C.~A. Arg\"uelles, A.~Diaz, A.~Kheirandish, A.~Olivares-Del-Campo, I.~Safa, and A.~C. Vincent {\em Rev. Mod. Phys.} {\bf 93} (2021), no.~3 035007, [\href{http://arxiv.org/abs/1912.09486}{{\tt arXiv:1912.09486}}].

\bibitem{LHAASO:2023gne}
{\bf LHAASO} Collaboration, Z.~Cao et~al. {\em Phys. Rev. Lett.} {\bf 131} (2023), no.~15 151001, [\href{http://arxiv.org/abs/2305.05372}{{\tt arXiv:2305.05372}}].

\bibitem{TibetASgamma:2021tpz}
{\bf Tibet ASgamma} Collaboration, M.~Amenomori et~al. {\em Phys. Rev. Lett.} {\bf 126} (2021), no.~14 141101, [\href{http://arxiv.org/abs/2104.05181}{{\tt arXiv:2104.05181}}].

\bibitem{KASCADEGrande:2017vwf}
{\bf KASCADE Grande} Collaboration, W.~D. Apel et~al. {\em Astrophys. J.} {\bf 848} (2017), no.~1 1, [\href{http://arxiv.org/abs/1710.02889}{{\tt arXiv:1710.02889}}].

\bibitem{PierreAuger:2022uwd}
{\bf Pierre Auger} Collaboration, P.~Abreu et~al. {\em Astrophys. J.} {\bf 933} (2022), no.~2 125, [\href{http://arxiv.org/abs/2205.14864}{{\tt arXiv:2205.14864}}].

\bibitem{Savina:2021cva}
{\bf Pierre Auger} Collaboration, P.~Savina {\em PoS} {\bf ICRC2021} (2021) 373.

\bibitem{PierreAuger:2022aty}
{\bf Pierre Auger} Collaboration, P.~Abreu et~al. {\em JCAP} {\bf 05} (2023) 021, [\href{http://arxiv.org/abs/2209.05926}{{\tt arXiv:2209.05926}}].

\bibitem{Keith:2020jww}
C.~Keith, D.~Hooper, N.~Blinov, and S.~D. McDermott {\em Phys. Rev. D} {\bf 102} (2020), no.~10 103512, [\href{http://arxiv.org/abs/2006.03608}{{\tt arXiv:2006.03608}}].

\bibitem{Slatyer:2015jla}
T.~R. Slatyer {\em Phys. Rev. D} {\bf 93} (2016), no.~2 023527, [\href{http://arxiv.org/abs/1506.03811}{{\tt arXiv:1506.03811}}].

\bibitem{Slatyer:2015kla}
T.~R. Slatyer {\em Phys. Rev. D} {\bf 93} (2016), no.~2 023521, [\href{http://arxiv.org/abs/1506.03812}{{\tt arXiv:1506.03812}}].

\bibitem{DAmico:2018sxd}
G.~D'Amico, P.~Panci, and A.~Strumia {\em Phys. Rev. Lett.} {\bf 121} (2018), no.~1 011103, [\href{http://arxiv.org/abs/1803.03629}{{\tt arXiv:1803.03629}}].

\bibitem{Giesen:2012rp}
G.~Giesen, J.~Lesgourgues, B.~Audren, and Y.~Ali-Haimoud {\em JCAP} {\bf 12} (2012) 008, [\href{http://arxiv.org/abs/1209.0247}{{\tt arXiv:1209.0247}}].

\bibitem{Slatyer:2016qyl}
T.~R. Slatyer and C.-L. Wu {\em Phys. Rev. D} {\bf 95} (2017), no.~2 023010, [\href{http://arxiv.org/abs/1610.06933}{{\tt arXiv:1610.06933}}].

\bibitem{Finkbeiner_2012}
D.~P. Finkbeiner, S.~Galli, T.~Lin, and T.~R. Slatyer {\em Physical Review D} {\bf 85} (Feb., 2012).

\end{thebibliography}\endgroup

\end{document}